\documentclass{aa}  
\usepackage{graphicx}
\usepackage{subfigure}
\usepackage{float}
\usepackage{placeins}
\usepackage{txfonts}

\begin{document}

   \title{Star-forming compact groups: Tracing the early evolutionary stages of compact group environments}
   
   \author{Ortiz-Gómez S.
          \inst{1},
          Torres-Flores S.
          \inst{1},
           Monachesi A.
          \inst{1},
          Montaguth G.P.
          \inst{2},
          Véliz Astudillo S.
          \inst{1},
          Mendes de Oliveira C.
          \inst{2},
          Olave-Rojas D. E.
          \inst{3},
          Lima-Dias C.
          \inst{1},
          Demarco R.
          \inst{4},
          Pallero D.
          \inst{5},
          Lopes A.R.
          \inst{6},
          Cortesi A.,
          \inst{7, 8}
          Telles E.
          \inst{9},
          Kanaan A.
          \inst{10},
          Ribeiro T.
          \inst{11},
          \and
          Schoenell W.
          \inst{12}
          }

   \institute{Departamento de Astronomía, Universidad de La Serena,
              Av. Raúl Bitrán 1305, La Serena, Chile\\
              \email{sebastian.ortiz@userena.cl}
         \and
             Instituto de Astronomia, Geof\'isica e Ci\^encias Atmosf\'ericas da Universidade de S\~ao Paulo, Cidade Universit\'ria, CEP:05508-900, S\~ao Paulo, SP, Brazil
         \and
            Departamento de Tecnologías Industriales, Facultad de Ingeniería, Universidad de Talca, Los Niches km 1, Curicó, Chile
         \and
            Facultad de Ciencias Exactas, Institute of Astrophysics, Universidad Andrés Bello, Sede Concepción, Talcahuano 7100, Concepción 4260000, Chile
        \and
            Departamento de Física, Universidad Técnica Federico Santa María, Avenida España 1600, Valparaíso, Chile
        \and
            Instituto de Astrofísica de La Plata, CONICET-UNLP, Paseo del Bosque s/n, B1900FWA, Argentina
        \and
            Observatório do Valongo/UFRJ, Ladeira do Pedro Antônio, 43 Centro, Rio de Janeiro, RJ 20080-090, Brazil
        \and
            Institute of Physics, Federal University of Rio de Janeiro, Av. Athos da Silveira Ramos 149, Rio de Janeiro, RJ 21941909, Brazil
        \and
            Observatório Nacional, Rua General José Cristino, 77, São Cristóvão, 20921-400, Rio de Janeiro, RJ, Brazil
        \and
            Departamento de Física, Universidade Federal de Santa Catarina, Florianópolis, SC, 88040-900, Brazil
        \and
            NOAO, 950 North Cherry Ave. Tucson, AZ 85719, United States
        \and
            The Observatories of the Carnegie Institution for Science, 813 Santa Barbara St, Pasadena, CA 91101, USA
}            

   \date{Received 25 July, 2025; accepted 13 January, 2026}
 
  \abstract
{In the context of pre-processing — a scenario where galaxies quench their star formation within substructures before falling into clusters — we investigate the impact of environment on the physical and morphological properties of galaxies in compact groups (CGs), focusing specifically on a sample of star-forming CGs (SFCGs).}
{Our aim is to characterize the physical and morphological properties of galaxies in SFCGs — analogs to the Blue Infalling Group — and understand how the environment influences their evolution.}
{We used photometric techniques to derive stellar masses and star formation rates (SFRs). Morphological parameters were extracted from DECaLS images, and we obtained parametric properties such as the Sérsic index ($n$) and effective radius ($R_e$) using GALFITM and non-parametric indices — including Gini, $M_{20}$, and asymmetry — from the same data. These indicators allowed us to classify galaxies into E/S0/Sa, Sb/Sc/Ir, and merger types. All measurements were compared to a control sample of field galaxies to assess environmental effects.}
{We find no significant differences in $n$ and $R_e$ between SFCG and field galaxies, in contrast to results for other CG samples. However, SFCG galaxies exhibit higher specific SFRs than field counterparts. About 16\% of SFCG galaxies show merger features and elevated asymmetry. These mergers also present enhanced SFRs compared to both other SFCG types and the field.}
{We propose that SFCGs represent an earlier evolutionary phase of CGs, supported by their lower velocity dispersions and moderate crossing times in addition to the observed SFR enhancement and absence of pronounced morphological transformation. Galaxy mergers in this phase appear to enhance, rather than suppress, star formation.}

\authorrunning{Ortiz-Gómez et al.}
\titlerunning{SFCGs: Tracing the early evolutionary stages of CG environments}
   \maketitle

\section{Introduction}\label{sec:intro}

The environment in which galaxies reside plays a key role in the cessation of star formation, driving the so-called environmental quenching \citep{Kormendy_2004, Peng_2010}. A major mechanism influencing galaxy evolution is gravitational interaction, where mergers \citep{Toomre_1972, Shlosman_2013} can strongly alter morphology and gas content, particularly when the mass ratio exceeds 1:3 \citep{Hopkins_2006, Puech_2019}. Such events are more common in galaxy groups, where the low velocity dispersion (100–300\,km\,s$^{-1}$; \citealt{Li_2012}) favors mergers. Additionally, tidal stripping can remove gas and stars through gravitational forces, modifying morphology and star formation capacity \citep{Moore_1998, Boselli_2014, Darvish_2024}.

Galaxies can also interact hydrodynamically with the gas in the intracluster medium (ICM) as they move through it inside the cluster. Gas can be removed from a galaxy and mix with the hot ICM, a process called ram-pressure stripping (RPS; \citealt{Gunn_and_Gott_1972, Domainko_2006, Jaffe_2018}). In some cases, as the gas is being removed, it compresses in the direction of the movement, triggering star formation as the galaxy infalls \citep{Ge_2023}. These hydrodynamical effects were believed to exclusively affect galaxies in clusters since it is most common that these structures reach the density of the ICM and the high velocity of the infalling galaxy, which increases the RPS effect. However, recent studies have shown that this phenomenon also occurs in lower mass structures, such as groups, where galaxies could also be losing gas through RPS \citep{Salem_2015, Kolcu_2022}. This has also been observed in simulations \citep{Simpson_2018, Pallero_2022}.

Using analytical models, \cite{Fujita_2004} proposed that galaxies may begin quenching before experiencing RPS in clusters, likely due to environmental effects in their host substructures. \cite{Haines_2015} showed that the fraction of star-forming galaxies rises beyond 2$r_{200}$ in clusters but remains below the field level up to 3$r_{200}$. Similarly, \cite{Lopes_2024} found that galaxy groups in infall regions already exhibit lower star-forming fractions than field galaxies (FGs). \cite{Morgan_2024} also recently found significant truncation in the star-forming disks outside the virial radius of the Virgo cluster, with a population of truncated disks too high to be produced solely by RPS, suggesting an environmental effect prior to cluster infall. This evidence supports the ``pre-processing'' scenario, where quenching begins in smaller structures such as groups prior to cluster infall, highlighting the critical role of the environment in galaxy evolution, which has also been found in simulations \citep{Pallero_2019}.

In this context, compact groups of galaxies (CGs) can be ideal laboratories to study the effects of the environment on galaxy evolution since their population, compactness, and dynamical properties enhance different phenomena such as mergers and tidal effects \citep{Hickson_1988, MdeOliv_1994, Alatalo_2015}. CGs usually show a higher fraction of early-type galaxies \citep[ETGs;][]{Coziol_2007, Walker_2010} and are slightly more luminous, more compact, and concentrated than those in the field \citep{Coenda_2012, Montaguth_2023}.

Concerning galaxy morphology in CGs, \citet{Montaguth_2023} reported that transition galaxies (those with a low Sérsic index and red colors) exhibit a bimodal Sérsic index distribution \citep{Sersic_1963}, indicating ongoing morphological transformation. They also found a higher quenched fraction and lower median specific star formation rates (sSFRs) in CGs than in the field, evidencing environmental quenching consistent with previous studies \citep[e.g.,][]{Alatalo_2015, Bitsakis_2016, Bianconi_2018}. These works suggest that star formation suppression arises not only from gas depletion but also from turbulence induced by shocks. Furthermore, \citet{Montaguth_2024} showed that 27\% of CGs belong to larger structures, where morphological and physical transformations are enhanced. As these studies focus on Hickson-like CGs, investigating less evolved systems is essential to understanding environmental effects in earlier evolutionary stages.

In the context of pre-processing, \cite{Cortese_2006} studied the Blue Infalling Group (BIG), a CG falling toward the cluster Abell 1367. Here we witness for the first time in the local Universe a CG infalling into the core of a dynamically young cluster. Galaxies in this CG suffer from physical and morphological transformations induced by the environment they are immersed in. In this context, the BIG would be an example of a galaxy being pre-processed in the local Universe, meaning local environmental processes take place in order to change the properties of galaxies
before their accretion into bigger structures. Although the BIG has been deeply studied, there are several open questions regarding these kinds of systems, including whether these types of events are a normal phenomena, the role that these SFCGs play in the evolutionary sequence of CGs, and whether they are a missing link in the evolution of galaxies in the hierarchical scenario of structure formation.

In this paper we aim to determine the properties of galaxies in SFCGs analogous to the BIG and understand how the environment affects the evolution of star-forming galaxies in young galaxy groups regarding their physical and morphological properties. We developed our analysis over a sample of 280 SFCGs taken from \cite{Hernandez-Fernandez_2015}, and we used deep photometric images and catalog information from the Dark Energy Camera Legacy Survey (DECaLS; \citealt{Dey_2019}).

This paper is organized as follows: In Sect. 2 we describe the data, the sample, and the main characteristics of the images, surveys, and catalogs. In Sect. 3 we describe the method followed to obtain the main morphological and physical parameters of galaxies, the steps, the software we utilized, and the information obtained from other catalogs. In Sect. 4 we present the results and compare them with the control sample data. In Sect. 5 we discuss the evolutionary stage of the SFCGs and the repercussions it may have in galaxy evolution. Finally, Sect. 6 is dedicated to the main conclusions of this paper. In this work we assume a flat cosmology, with $H_{0}$ = 67.4 $km s^{-1} Mpc^{-1}$, $\Omega_m$ = 0.315, and $\Omega_{\lambda}$ = 0.685 \citep{Planck_2020}. 

\section{Sample and data}\label{sec:data}

\subsection{The star-forming compact group sample}\label{sec:sfcg_sample}

The sample used in this work was selected by \cite{Hernandez-Fernandez_2015}, whose aim was to identify CGs composed mainly of star-forming galaxies in the local Universe that can serve as BIG analogs. They employed the GALEX All-sky Imaging Survey (AIS; \citealp{Martin_2005}), selecting sources with far-UV (FUV) magnitudes of $17 < \mathrm{FUV} < 20.5$, similar to BIG galaxies, since the FUV band traces young stellar populations with shorter lifetimes than those probed by NUV \citep{Martin_2005, Haines_2008}. A color criterion of $-1.50 < (FUV-NUV)_d < 2.75$ was applied, and magnitudes corrected for foreground extinction using the Cardelli law \citep{Cardelli_1989}. To reduce stellar contamination, the search was limited to sources located more than 15 degrees from the Galactic plane.

The search for CGs made by the authors was done by applying a friends-of-friends algorithm over the sample of 925428 UV sources in the space of celestial coordinates, imposing a maximum linking length of 1.5 arcmin, which corresponds to a projected distance of {$\sim$}88 kpc at z=0.05. After crossmatching the groups obtained with galaxy catalogs compiled by the NASA/IPAC Extragalactic Database (NED), they applied additional criteria, such as considering as groups those systems with at least three sources compiled as ``galaxy" by NED and/or at least two galaxies with accordant redshifts. \cite{Hernandez-Fernandez_2015} produced a catalog of 280 SFCGs up to z {$<$} 0.17, composed of 226, 39, 11, and 4 groups of four, five, six, and seven bright UV members, respectively. Seventy-five percent of the SFCGs have at least one member with a spectroscopic redshift available from NED, and over 40{$\%$} of the SFCGs have redshifts measured for two or more galaxies.

The authors compared some physical properties, including dynamics of groups (those with some members with redshift available), with other group catalogs, such as the optically selected Hickson Compact Groups sample (HCG; \citealt{Hickson_1982}), the Near Infrared (NIR)-selected 2MASS Compact Group sample (2MCG; \citealt{Diaz-Gimenez_2012}), and the catalog A of compact groups selected in the Sloan Digital Sky Survey (SCGA; \citealt{McConnachie_2009}). They found that the SFCGs have lower velocity dispersions ({$\sigma_{l-o-s} \sim 120$}km/s), small crossing times ({$H_{0}t_{c} \sim$} 0.05), and high star formation content (95{$\%$} of star-forming galaxies).

In this work, we analyze a final sample of 207 SFCGs for the morphological analysis (see Sects. \ref{sec:optical-images-and-catalogs} and \ref{sec:meth_morph_param} for details on the final selection) and 110 SFCGs for the physical analysis (see Sect. \ref{sec:stellar_masses} for details on this final selection). We assessed the potential presence of non–UV-bright members in our SFCG sample, following \citet{Hernandez-Fernandez_2015}, who found 59 of the 280 SFCGs to have counterparts within 2.5 arcmin. Our visual inspection revealed eight groups (out of 207) with one to two apparently quiescent galaxies along the line of sight and nine with similarly close galaxies showing colors consistent with star-forming members. Their membership remains uncertain due to incomplete redshift data. Overall, these cases are rare, and the sample is still dominated by star-forming galaxies. We emphasize that this work focuses on the morphological and physical properties of such galaxies in dense environments.

\subsection{Optical images and catalogs: DESI Legacy Imaging Surveys and S-PLUS}\label{sec:optical-images-and-catalogs}

In this work we use data from DECaLS (\citealt{Dey_2019}) from the DESI Legacy Imaging Surveys. DECaLS provides optical imaging for two-thirds of the DESI footprint, covering both the north Galactic cap region at Dec {$\gtrsim$} +32° and the south Galactic cap region at Dec {$\lesssim$} +34°. Due to the combination of a large field of view and a high sensitivity from 400 to 1000 nm, DECam is a very efficient option for obtaining photometry in the \textit{g, r}, and \textit{z} bands, with the DR10 also adding the \textit{i} band. These characteristics make it ideal to determine morphological properties of galaxies.

Apart from images, DECaLS also provides catalogs with photometric data extracted using \texttt{The Tractor} package \citep{Lang_2016}, which uses a probabilistic method to model sources in an image, classifying them as point sources (PSFs), round exponential galaxies with a variable radius, de Vaucouleurs profiles (elliptical galaxies), exponential profiles (spiral galaxies), and Sérsic profiles. Other information about photometry, calibrations, source detection, and other parameters is available in \cite{Dey_2019}. The predicted Legacy Survey depths for two observations at 1.5" seeing were \textit{g}=24.7, \textit{r}=23.9, \textit{i}=23.0, and \textit{z}=23.0, reaching a surface brightness limit in the \textit{r} band of 27.9 mag/arcsec{$^2$} \citep{Hood_2018}.

In this work we downloaded the Legacy Survey DR10 catalogs from the NOIRLab AstroDataLab portal.\footnote{\url{https://datalab.noirlab.edu}} We performed a cone search around the center coordinates of each group, with a 6 arcmin radius, and then we made a crossmatch with the SFCG galaxy catalog. From the 280 SFCGs, 226 were in the DECaLS footprint containing 970 members. By visually inspecting the groups, we noticed that there were interlopers, such as stars in the SFCG catalog, which were filtered using the "PSF" type in the Legacy Survey catalog. We found 135 stars among all groups, and they were removed, leaving 835 galaxies in total. On average, each group is composed of approximately four galaxies. 

We note that previous studies used multiwavelength data to perform morphological analyses. In particular, \cite{Montaguth_2023} analyzed a sample of CGs by using data from the Southern Photometric Local Universe Survey (S-PLUS; \citealt{Mendesdeoliveira_2019}), which contains 12 filters and uses a 0.8-mt telescope at Cerro Tololo. This survey has a depth of \textit{r} = 21.18 mag at S/N > 3 \citep{Herpich2024}. Details of S-PLUS can be found in \cite{Mendesdeoliveira_2019}. In order to explore possible biases when comparing results arising from the DECaLS and S-PLUS data (and to make a fair comparison), in Sect. \ref{sec:splus_decals} and Appendix \ref{sec:app_morph-param-spdec} we compare morphological parameters derived from both surveys for a subsample of the SFCGs.

\subsection{The control sample}\label{sec:the_control_sample}

To be able to extract conclusions regarding the effect of the environment on the SFCG galaxies, it is essential to define a control sample containing FGs. For this, we had to constrain galaxies in the control sample so that they would have similar properties as the galaxies in the SFCG sample, including galaxy mass and redshift.

In this work, we selected the control sample galaxies from the \cite{Yang_2007} catalog (available on the official website\footnote{\url{https://gax.sjtu.edu.cn/data/Group.html}}), which contains galaxy groups from the Sloan-Digital Sky Survey (SDSS DR4; \citealt{Adelman_2006}) selected using a halo-based group finder. There are groups where the number of members is N = 1, i.e., isolated galaxies, which we selected to conform the control sample.

We applied several selection criteria to construct the final control sample. Field galaxies were first selected within the DECaLS footprint to ensure a consistent analysis with SFCG galaxies, and they were further restricted to those included in the GALEX–WISE–Sloan–Legacy Catalog (GWSLC; \citealt{Salim_2018}), which provides physical properties based on spectral energy distribution (SED) fitting that are compared with our derived measurements in Sect. \ref{sec:stellar_masses}. From this catalog, 2000 galaxies were randomly drawn within the redshift range $0.01 < z < 0.17$, and 1200 were selected to match the stellar mass distribution of the SFCGs (see Sect. \ref{sec:stellar_masses}). Applying the same color cut used for the SFCG selection yielded a final control sample of 743 FGs. We checked the redshift distribution of both samples, and although we found a slight difference in one of the redshift bins, it does not affect the morphological results.

\section{Methodology}\label{sec:methodology}

\subsection{Morphological parameters: Parametric approach}\label{sec:meth_morph_param}

To extract morphological parameters from the DECaLS images, we used a modified version of ``MorphoPLUS"\footnote{\url{https://github.com/GMontaguth/MorphoPLUS}} (\citeauthor{Montaguth_2025_B} submitted), a Bash/Python pipeline that employs SExtractor \citep{Bertin} and the MegaMorph code \citep{Hussler_2013}. MegaMorph performs multiwavelength 2D modeling using GALFITM, an extension of GALFIT \citep{Peng_2002, Peng_2010_galfitm}, capable of fitting Sérsic, exponential, de Vaucouleurs, Nuker, Gaussian, and Moffat profiles. GALFITM models the wavelength dependence of structural parameters via Chebyshev polynomials, enabling simultaneous fitting across bands. The best fit is obtained by minimizing $\chi^2$ using the Levenberg–Marquardt algorithm. Detailed procedures and fitting diagnostics are provided in Appendix \ref{sec:appendix_galfitm}.

We adopted a single-component Sérsic profile \citep{Sersic_1963}, where the Sérsic index \textit{n} describes the curvature of the light profile and the effective radius $R_e$ encloses half the total luminosity. For \textit{n} = 4, the profile reduces to the de Vaucouleurs law, which is typical of elliptical galaxies and bulges. For \textit{n} = 1, it becomes an exponential profile, which is characteristic of disks. Larger values of \textit{n} indicate more centrally concentrated light and higher surface brightness at large radii.

From the 226 groups, we were able to correctly obtain the morphological parameters for 640 galaxies in 207 SFCGs. The other SFCGs presented image problems since some of the images were incomplete, and it was not possible to obtain reliable morphological parameters (see an example of a bad fit in Fig. \ref{fig:bad_fit}). We selected the morphological parameters with reduced $\chi^2 < 2$. For completeness, in the morphological analysis we considered only 474 galaxies that are located in groups with at least one member with redshift available. The same procedure was followed to derive morphological properties for the 743 FGs.

\subsection{Morphological parameters: Non-parametric approach}\label{sec:meth_non_param}

To gain a more detailed understanding of the morphology of galaxies in the SFCGs, we obtained the non-parametric morphological metrics for galaxies in the sample. A non-parametric morphology refers to methods for characterizing the structure and shape of galaxies without assuming any specific functional form or model for their light distribution. These methods rely on statistical or geometric measurements of the galaxy's appearance derived directly from its pixel intensity distribution in an image.

We employed several non-parametric indicators to describe galaxy morphology, such as asymmetry \citep{Schade_1995, Abraham_1996, Conselice_2003}, Gini \citep{Abraham_2003, Lotz_2004}, and $M_{20}$ \citep{Lotz_2004}, as defined in \cite{Sazonova}. These parameters enable the classification of galaxies into E/S0/Sa, Sb/Sc/Ir, or mergers. Morphological measurements were obtained using the \texttt{Astromorphlib} library\footnote{\url{https://gitlab.com/joseaher/astromorphlib}} \citep{hernandez_jimenez_2022, Krabbe_2024} and the \texttt{statmorph} package,\footnote{\url{https://github.com/vrodgom/statmorph}} both of which provide automated pipelines from image processing to catalog generation. Non-parametric parameters were extracted in the $r$ band for 678 SFCG and 743 FGs, considering only those with \texttt{flagstatmorph} = 0 or 1 to ensure reliability. We did not exclude the $\chi^{2} > 2$ galaxies in the non-parametric analysis.

\subsection{Stellar masses}\label{sec:stellar_masses}

To obtain the mass of the stellar content of the galaxies, we used the calibration proposed by \citealt{Taylor_2011}(see Eq. 8 therein). These authors used (\textit{g - i}) rest-frame color and the absolute magnitude in the \textit{i} band to compute the stellar mass ($M_*$), which comes from the observed relation between (\textit{g - i}) and $M_*$/$L_i$, with $L_i$ as the luminosity on the \textit{i} band, assuming a Chabrier IMF \citep{Chabrier_2003}.

It is important to note that since we do not have a redshift confirmation for every galaxy in the groups, for this calculation we considered only those SFCGs with at least one galaxy member with a confirmed redshift, and we assumed that redshift as the redshift of the group. Also, the \textit{i} filter was only added in the last data release of DECaLS, so there are several SFCGs that do not have images available for that filter. Therefore, we calculated the stellar masses for 378 galaxies in 110 SFCGs.

We verified the reliability of this method by comparing the stellar masses of the FGs calculated as explained above with those available in the GWSLC \citep{Salim_2018}. As mentioned in Sect. \ref{sec:the_control_sample}, the masses in this catalog were computed using SED fitting, considering data from mid-infrared to FUV. Figure \ref{fig:photvssed} shows that the correlation between masses obtained using the method published by \cite{Taylor_2011} and those from the GWSLC is practically linear, with a very high correlation coefficient. Thus, it is reliable to use stellar masses obtained through the color relation by \cite{Taylor_2011}.

\begin{figure}[h!]
    \centering
    \includegraphics[width=1\linewidth]{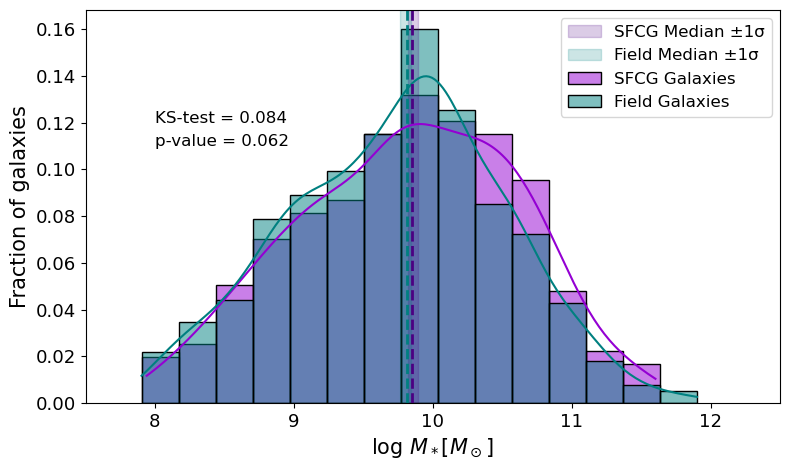}
    \caption{Mass distribution of the galaxies in the SFCGs and FGs. The y-axis shows the fraction of galaxies with respect to the total of each sample. Dashed lines represent the median of each distribution. In violet are the SFCGs, and the FGs are in teal. The shaded regions are 1$\sigma$ uncertainties around the median value of the distributions, obtained using bootstrapping. The violet and teal curves show the KDE for the distributions.}
    \label{fig:mass-distribution-control-groups}
\end{figure}

In Fig. \ref{fig:mass-distribution-control-groups} we show the stellar mass distribution of the galaxies in the SFCGs and FGs. We note that we removed 22 galaxies in the SFCG sample with stellar masses lower than $10^{7.9} M_{\odot}$ that did not have an FG counterpart so that both mass distributions of the samples analyzed agree. We thus ended up with a final number of 356 galaxies for the physical analysis in the SFCGs. The mass of the galaxies in the SFCGs ranges between {$10^{7.94} M_{\odot}$} and {$10^{11.96} M_{\odot}$}, with a peak at {$10^{9.85} M_{\odot}$}. As it is possible to see, there is a considerable amount of galaxies in lower mass bins ($M_{*} < 10^{9}$). This can be explained due to the selection criteria since the SFCGs were selected mainly by the UV emission of their members, and they should be related to dwarf star-forming galaxies. We emphasize that we cannot discard the existence of tidal dwarf galaxies in our sample.

We performed a Kolmogorov-Smirnov (KS) test using the \texttt{scipy} Python library to check if both mass distributions could be comparable. The p-value of the KS test is 0.06, which is higher than 0.05 and thus indicative that the samples have a similar mass distribution.

\subsection{Star formation rates}\label{sec:star_formation_rate}

Star formation rates (SFRs) were estimated using the calibrations proposed by \citet[their Eq. 3]{Iglesias_2006}, which consider that the SFR can be computed using the NUV luminosity corrected for the light absorption by dust. In their work, the authors used a sample of galaxies selected using GALEX \citep{Martin_2005} and the IR information at 60 {$\mu$}m from IRAS \citep{Moshir_1990} to correct the UV luminosity. The DECaLS catalogs provide data from the Wide-field Infrared Survey Explorer (WISE; \citealt{Wise_2010}), in the \textit{W1, W2, W3}, and \textit{W4} bands (3.4, 4.6, 12, and 22 {$\mu$}m, respectively). The dust heated by star formation is better traced in the mid-infrared wavelength range, from 10 to 25-40 microns. Therefore, we used the 22 {$\mu$}m information to correct the luminosity in the UV using Eq. (5) in Table (3) from \cite{Hao_2011}.

With luminosities corrected for dust and using {$L_{\odot, bol} = 3.86\times10^{33}$} erg/s, we computed the SFR for the SFCGs and FGs. \cite{Iglesias_2006} showed that the SFR$_{NUV}$ is slightly higher than the SFR computed with other methods, but the difference is always within 15\%. In Fig. \ref{fig:sfr-correlation} we show a comparison between the SFR of our FGs obtained through this method versus the SFR from the GWSLC obtained using SED-fitting techniques, where that difference is also observed. Since we used the same method for both samples and the aim is to compare how this property changes with respect to the environment, we adopted the SFR$_{NUV}$ estimates, which can be obtained for the SFCGs and FGs, given that the SFCG galaxies are not in the GWSLC. However, we note that their absolute values are likely not correct.

\section{Results}\label{sec:results}
\subsection{Galaxy classification in the star-forming compact group sample}\label{sec:gal_classification}

Using MegaMorph and multiwavelength data, \cite{Vika_2015} classified galaxies of different morphological types according to their \textit{n} index and (\textit{u - r}) color. DECaLS does not provide information in the \textit{u} band, which is why we used a (\textit{g - r}) color limit value. The determination of the (\textit{g - r}) color limit is explained in Appendix \ref{sec:appendix_classification}. The Sérsic index and (\textit{g - r}) values on which regions are defined are $n_r < 2.5$ and $(g - r) < 0.67$ for late-type galaxies (LTGs), while $n_r > 2.5$ and $(g - r) > 0.67$ define ETGs. Regions with intermediate properties are established as the ``transition region," defined by disk galaxies ($n_r < 2.5$) with redder colors $(g - r) > 0.67$, and the ``other region," which includes spheroidal galaxies ($n_r > 2.5$) with bluer colors, i.e., $(g - r) < 0.67$ \citep{Montaguth_2023}. 

In this way, $n_r$ can be used to separate early- and late-type structures, while color is used as a proxy of the star formation rate of a galaxy. Here, galaxies with a low $n_r$ and bluer color are LTGs undergoing star formation and are different from galaxies in the transition region, which are galaxies with a late-type structure and lower SFR. Galaxies with a higher $n_r$ and redder color resemble ETGs with a lower SFR than in the other region, where we find galaxies with an early-type structure with active star formation. For simplicity, in this work we use the definition of ETG and LTG from the Sérsic color plane following Vika’s classification, but we note that ETGs and LTGs in this work do not only refer to structural properties.

\begin{figure}[h!]
    \centering
    \begin{subfigure}
        \centering
        \includegraphics[width=1\linewidth]{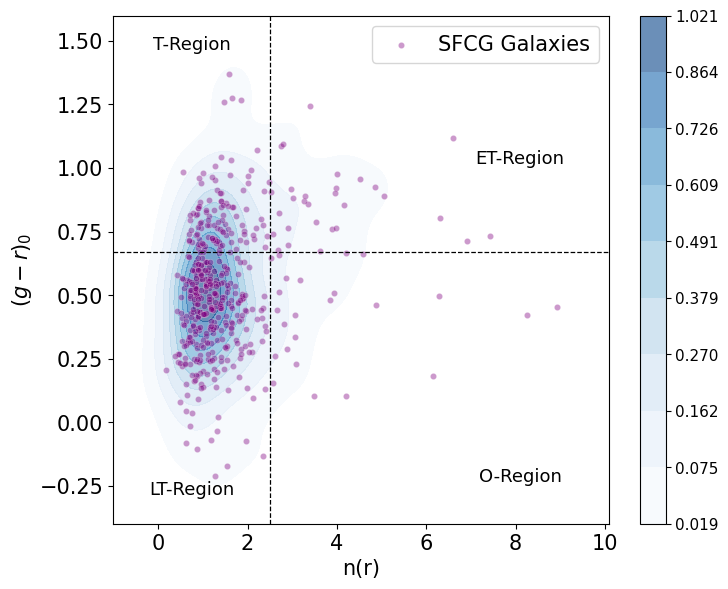}
    \end{subfigure}
    \hfill
    \begin{subfigure}
        \centering
        \includegraphics[width=1\linewidth]{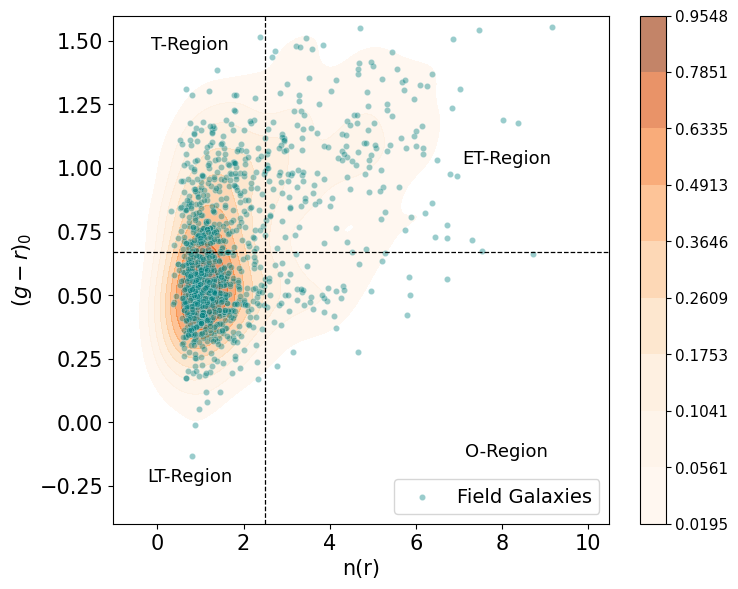}
    \end{subfigure}
    \caption{Galaxy classification according to their morphology and color. The vertical line is at $n_r$ = 2.5, while the horizontal line is at (\textit{g - r}) = 0.67. Top panel: Galaxies in the SFCGs. Lower panel: FGs. The color bar shows the KDE.}
    \label{fig:regions}
\end{figure}

Figure \ref{fig:regions} shows the distribution of galaxies in the Sérsic color plane for SFCGs (top) and FGs (bottom), with contours representing the 2D kernel density estimation (KDE). In SFCGs, galaxies mainly occupy the late-type region, followed by the transition region, with fewer objects in the early-type and other regions. As summarized in Table \ref{tab:regions-fraction-table}, LTGs comprise $\sim$65\% of SFCG galaxies, while in the field they represent 61\%. The field distribution also peaks in the late-type region, extending toward the transition and early-type regions, as expected since isolated galaxies, predominantly LTGs, evolve through slower secular processes \citep{Kormendy_2004, Peng_2010}. The ETG fraction is slightly higher in the field than in SFCGs (11\% and 6\%, respectively).

The transition region displays the same relative fraction for galaxies in both environments. This region is of particular interest since the galaxies in it may be experiencing a physical change regarding their evolution, as has been observed in the analysis of more evolved CGs \citep{Montaguth_2023}. In the "other region," we find a very small sample since it is not that common to find blue spheroid-shaped galaxies at low z, and 30\% of these galaxies have stellar masses associated with the dwarf kind.

\begin{table}[h!]
\caption{Relative fraction (amount) of galaxies of each type.}
\centering
\begin{tabular}{|c|c|c|}
\hline
\textbf{Galaxy Type} & \textbf{SFCG} & \textbf{Field} \\ \hline
Early Type & $6.34 \pm 1.19$\% (30) & $10.6 \pm 1.14$\% (79) \\ \hline
Late Type  & $65.54 \pm 2.16$\% (310) & $60.81 \pm 1.74$\% (453)\\ \hline
Transition  & $23.04 \pm 1.91$\% (109) & $22.68 \pm 1.54$\% (169) \\ \hline
Other  & $5.07 \pm 1.01$\% (25) & $5.64 \pm 0.85$\% (42) \\ \hline
\end{tabular}
\tablefoot{Galaxy types are defined by regions in Fig. \ref{fig:regions}. Uncertainties were estimated using bootstrapping with a 68\% confidence interval.}
\label{tab:regions-fraction-table}
\end{table}

\subsection{Morphological properties of SFCG galaxies}\label{sec:morph_properties}

\begin{figure}[h!]
    \centering
    \includegraphics[width=0.7\linewidth]{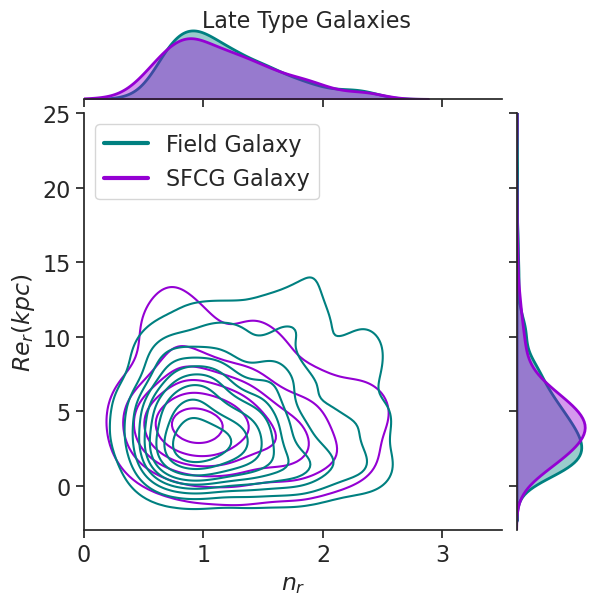}
    \caption{Effective radius vs. Sérsic index in the \textit{r} filter for the LTGs in the SFCGs (violet) and in the field (teal).}
    \label{fig:late_n_re}
\end{figure}

\begin{figure}[h!]
    \centering
    \includegraphics[width=0.7\linewidth]{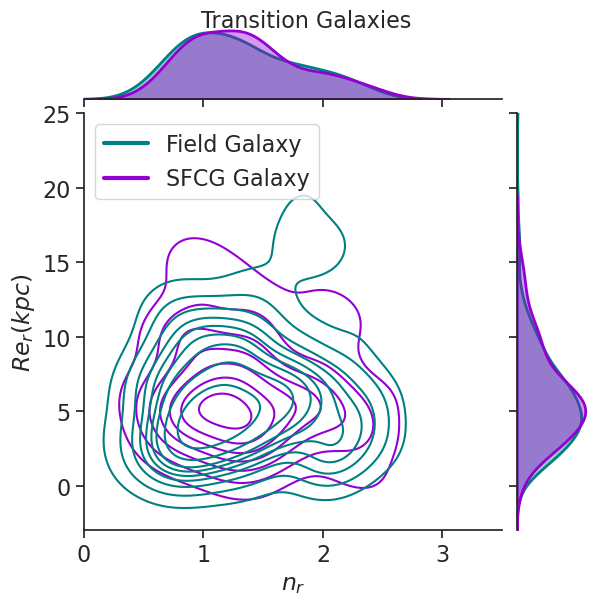}
    \caption{Effective radius vs. Sérsic index in the \textit{r} filter for the transition galaxies in the SFCGs (violet) and in the field (teal).}
    \label{fig:trans_n_re}
\end{figure}

In order to analyze the morphological properties of galaxies in the SFCGs and the field, we compared $R_e$ and \textit{n} in both environments. Figures \ref{fig:late_n_re} and \ref{fig:trans_n_re} show the KDE contours of each distribution for LTGs and transition galaxies, respectively, to facilitate analysis of where differences exist according to their type. We use $R_e$ in kiloparsec, so we only considered groups with at least one group member with redshift available, assuming it is the redshift of the group. Figs. \ref{fig:late_n_re} and \ref{fig:trans_n_re} present our results in the \textit{r} band (hence $Re_r$ and $n_r$). Figures \ref{fig:late_n_re_all} and \ref{fig:trans_n_re_all} show our results in the \textit{g} and \textit{z} bands. 

From Fig. \ref{fig:late_n_re} we observed that \textit{n} distributes mostly in a very narrow range below \textit{$n_r \approx 1.5$}. We observed that LTGs present a more concentrated distribution in the Sérsic index, with no significant differences between each environment. The distribution of $n$ peaks around $n_{r} = 1$ and decays toward higher values. Regarding transition galaxies, which are shown in Fig. \ref{fig:trans_n_re}, there is a slight difference in the Sérsic index with respect to LTGs. In this region, we observed redder galaxies, and in both environments there is a higher fraction of galaxies with larger $n_r$ values compared to the late-type population. The Sérsic index does not show a strong concentration around a single value, and we find that the fraction of galaxies with $n_r > 1.5$ is 10$\%$ higher in transition galaxies compared to LTGs.   

The $R_e$ values for LTGs are similar in both environments, with galaxies in the SFCGs presenting a slightly larger {$R_e$} than galaxies in the field. Although the difference is statistically significant, in practice the difference of the median $R_e$ for galaxies in both environments is 0.4 kpc. At the average redshift of the sample ($z \sim 0.05$), that difference is 0.4", which is below the average seeing at CTIO. Figure \ref{fig:trans_n_re} shows that the median $R_e$ is slightly higher for transition galaxies than for LTGs in SFCGs and FGs (see also Table \ref{tab:param_values_regions}). It also shows a wider Sérsic index distribution and a more evident increase of this parameter toward redder colors. We did not find any significant differences in the median effective radius of SFCGs and FGs in the transition region.

We did not find any clear bimodality in one or both parameters, as detected in \cite{Montaguth_2023}, for a more evolved CG sample. This suggests that if SFCG galaxies are undergoing morphological transformations, they are in a very early stage, which is why we cannot see very strong effects. Also, these findings suggest that morphological transformations in these environments occur after the effects on star formation.

\subsection{Multiple band fitting: DECaLS and S-PLUS morphological results}
\label{sec:splus_decals}

The structural parameters of galaxies can be strongly affected by the quality of the images and the filters used to obtain them. In Appendix \ref{sec:app_morph-param-spdec} we aim to analyze the behavior of morphological properties of galaxies in S-PLUS and DECaLS. S-PLUS contains 12 filters, which theoretically would be an advantage in GALFITM modeling, with the hope being that with a greater number of filters we will constrain morphological parameters with more accuracy. On the other hand, DECaLS only contains four broadband filters, but it is deeper and has better resolved images. 

Although the morphological parameters in S-PLUS are slightly higher, with a difference of $n = 0.11 \pm 0.01$ and $R_e = 0.48 \pm 0.09$ kpc higher than those in DECaLS, the uncertainties overlap, and the differences are insignificant. Therefore, these values are not sufficiently high to produce a different interpretation of the results. More details about this analysis are given in Appendix \ref{sec:app_morph-param-spdec}.

\subsection{Correlation with physical properties}\label{sec:corr_phy_prop}

\begin{figure}[h!]
    \centering
    \includegraphics[width=1\linewidth]{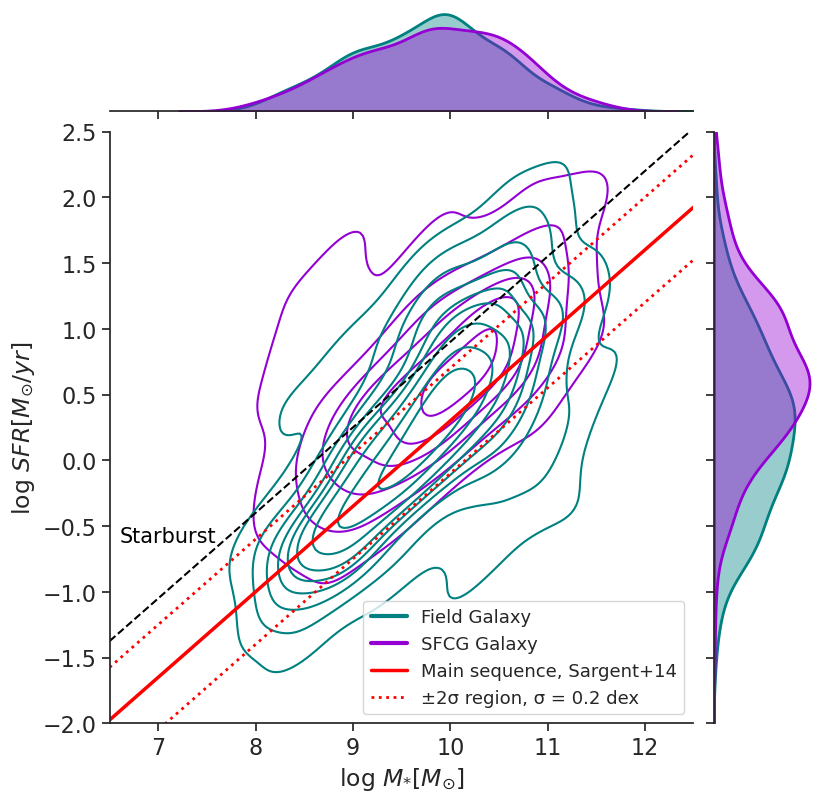}
    \caption{Star formation rate compared against stellar mass KDE contours for galaxies in the SFCGs (violet) and the field (teal). The dashed black line represents the division between starburst and no starburst galaxies \citep{Jarvis_2020}, and the red line represents the main sequence \citep{Sargent_2014}.}
    \label{fig:main-sequence}
\end{figure}

To analyze the general state of galaxies in the SFCGs with respect to galaxies in the field, we compared their stellar mass and SFR obtained in Sects. \ref{sec:stellar_masses} and \ref{sec:star_formation_rate}. Figure \ref{fig:main-sequence} shows the KDE contours for galaxies in both environments with their respective distributions in the marginal plots. In general, we find that galaxies in the SFCGs present a median log (SFR) = 0.41 $\pm$ 0.05 [$M_\odot/yr$] dex higher than that for galaxies in the field for galaxies at the same stellar mass. 

In Fig. \ref{fig:main-sequence} it can be seen that the majority of the FGs inhabit the star-forming main sequence (SFMS; \citealt{Sargent_2014}), with 64$\%$ of them located inside 2$\sigma$ of the main sequence. A fraction of these galaxies have a higher SFR, with 14$\%$ of them located in the starburst region. Also, 13$\%$ of the galaxies are located in the region of low star formation and high stellar masses, suggesting the presence of quiescent galaxies. 

On the other side, 84$\%$ of galaxies in the SFCGs are above the SFMS (red solid line), although 46$\%$ are inside 2$\sigma$ (dotted red lines) of it. Of the sample, 35$\%$ is in the starburst region (above the dashed black line) of the plane, while only 3$\%$ is located in the less active star-forming region. The center of the distribution is located between the main sequence and the starburst galaxies. The higher mass SFCG galaxies also present a relatively high SFR. The differences between the physical properties of the galaxies in each environment are a representation of the different populations that both environments contain, with a lack of ETGs in the SFCGs, as expected.

It is important to reiterate that the SFR may be slightly higher with the method we used in this work than the SED-fitting method (see Fig. \ref{fig:sfr-correlation}), so the SFMS may likely not be representative of these data nor of the starburst region. However, we remark that there is an intrinsic difference between SFCGs and FGs since they have been computed using the same method.

\begin{figure*}[h!]
    \centering
    \begin{subfigure}
        \centering
        \includegraphics[width=0.47\textwidth]{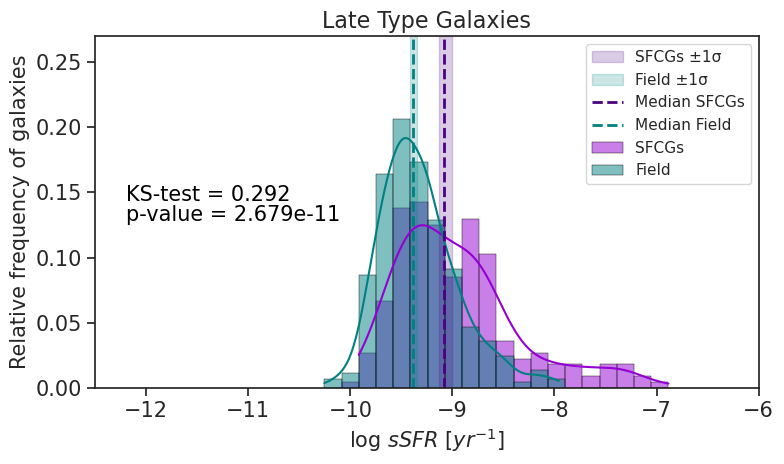}
    \end{subfigure}
    \begin{subfigure}
        \centering
    \includegraphics[width=0.47\textwidth]{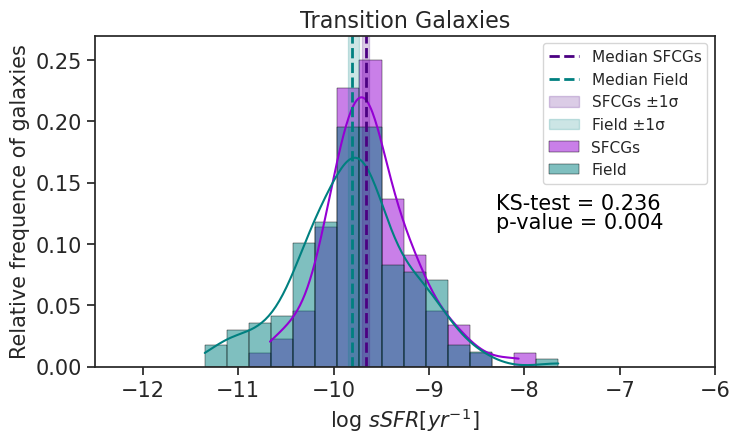}
    \end{subfigure}
    \caption{Specific star formation rate for galaxies in the SFCGs (violet) and in the field (teal). The left panel shows the distribution of the LTGs, and the right panel shows the distribution for transition galaxies. The dashed line shows the median of the distribution, and the shaded regions are the 1$\sigma$ uncertainty of the median value using bootstrapping.}
    \label{fig:ssfr_re_types}
\end{figure*}

Figure \ref{fig:ssfr_re_types} shows the distribution of log ($sSFR$) for galaxies in SFCGs and the field for LTGs (left panel) and for transition galaxies (right panel). The LTGs in the SFCGs show a broader distribution in log ($sSFR$) than those in the field, with a median 0.30 $\pm$ 0.07 dex higher. In the right panel of Fig. \ref{fig:ssfr_re_types}, we observed a broader distribution of the sSFR for transition galaxies in the field than for galaxies in the SFCGs since there is a larger part of the sample toward a lower sSFR. Based on the \cite{Wetzel_2013} criterion, which suggests that we should consider as quenched galaxies those that have a log ({$sSFR/yr^{-1}$}) {$< -11$}, we found a fraction of transition galaxies in the field already quenched, which is not found in SFCG galaxies. In this figure, it is clear that transition galaxies in the SFCGs are not yet quenched. In Table \ref{tab:phy_values_regions} we show the median values of the physical properties for each galaxy type.

\begin{table}[h!]
\caption{Physical properties for each galaxy type.}
\centering
\begin{tabular}{|c|c|}
\hline
\textbf{Galaxy Type} & \textbf{Median log (SFR)} \\ \hline
Early Type    & $0.85 \pm 0.20$ ($0.48 \pm 0.08$) \\ \hline
Late Type     & $0.55 \pm 0.05$ ($0.07 \pm 0.04$) \\ \hline
Transition    & $0.82 \pm 0.08$ ($0.50 \pm 0.06$) \\ \hline
Other         & $0.31 \pm 0.28$ ($0.33 \pm 0.16$) \\ \hline
Whole Sample  & $0.64 \pm 0.03$ ($0.23 \pm 0.03$) \\ \hline
\textbf{Galaxy Type} & \textbf{Median log (sSFR)} \\ \hline
Early Type    & $-9.91 \pm 0.21$ ($-10.14 \pm 0.10$) \\ \hline
Late Type     & $-9.08 \pm 0.06$ ($-9.38 \pm 0.03$) \\ \hline
Transition    & $-9.66 \pm 0.04$ ($-9.78 \pm 0.06$) \\ \hline
Other         & $-9.21 \pm 0.18$ ($-9.32 \pm 0.07$) \\ \hline
Whole Sample  & $-9.28 \pm 0.04$ ($-9.49 \pm 0.02$) \\ \hline
\textbf{Galaxy Type} & \textbf{Median $M_{*}$} \\ \hline
Early Type    & $10.75 \pm 0.11$ ($10.82 \pm 0.09$) \\ \hline
Late Type     & $9.60 \pm 0.09$ ($9.47 \pm 0.04$) \\ \hline
Transition    & $10.48 \pm 0.06$ ($10.30 \pm 0.05$) \\ \hline
Other         & $9.90 \pm 0.30$ ($9.79 \pm 0.09$) \\ \hline
Whole Sample  & $9.86 \pm 0.04$ ($9.82 \pm 0.04$) \\ \hline
\end{tabular}
\tablefoot{Galaxy types are defined by the regions in Sect. \ref{sec:gal_classification}. Values correspond to the SFCG galaxies, and in parenthesis are values for FGs. Uncertainties were estimated using bootstrapping with a 68\% confidence interval.}
\label{tab:phy_values_regions}
\end{table}

\subsection{Non-parametric morphological classification}\label{sec:non-parametric-approach}

\begin{figure*}[h!]
    \centering
    \begin{subfigure}
        \centering        \includegraphics[width=0.45\linewidth]{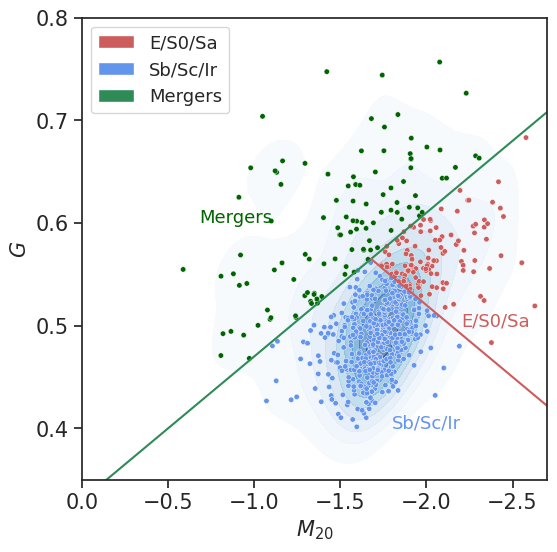}
    \end{subfigure}
    \hfill
    \begin{subfigure}
        \centering
    \includegraphics[width=0.45\linewidth]{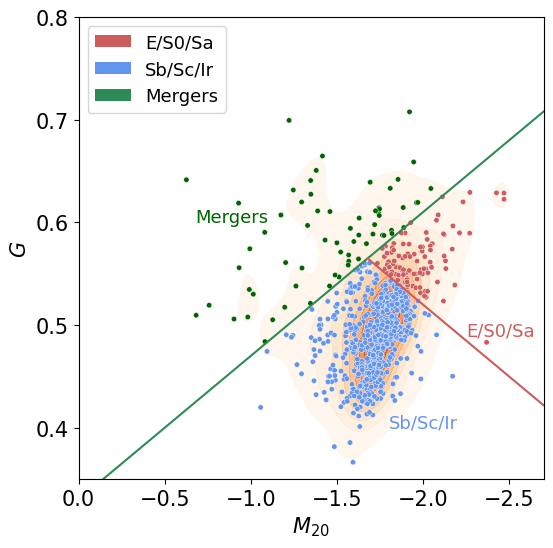}
    \end{subfigure}
    \caption{Galaxy classification according to their Gini and {$M_{20}$} indices. The blue, red, and green dots correspond to Sb/Sc/Ir, E/S0/Sa, and merger-type galaxies, respectively. The left panel corresponds to the SFCGs, while the right panel corresponds to the FGs. The G-$M_{20}$ plane and the KDE were built using 678 and 761 galaxies with reliable parameters from SFCGs and the field, respectively.}
    \label{fig:gini_m20_regions}
\end{figure*}

As mentioned in Sect. \ref{sec:meth_non_param}, we also performed a non-parametric morphological classification (NPC). Figure \ref{fig:gini_m20_regions} shows the Gini-$M_{20}$ plane \citep{Lotz_2004} following the \cite{Sazonova} scheme, distinguishing mergers, Sb/Sc/Ir, and E/S0/Sa galaxies for SFCGs (left) and FGs (right). Both samples cluster mainly in the Sb/Sc/Ir region, with SFCGs showing a broader spread toward mergers. As listed in Table \ref{tab:gini-m20-fraction-table}, Sb/Sc/Ir galaxies dominate (66\% in SFCGs, 78\% in FGs), while E/S0/Sa fractions are similar. This differs from the color-\textit{n} classification (Fig. \ref{fig:regions}), where only $\sim$6\% of SFCG galaxies were ETGs since the non-parametric method lacks color information and classifies red low-\textit{n} systems as disks.

\begin{table}[h!]
\caption{Relative fraction (amount) of galaxies for each Gini-{$M_{20}$} classification type.}
\centering
\begin{tabular}{|c|c|c|}
\hline
\textbf{Galaxy Type} & \textbf{SFCG} & \textbf{Field} \\ \hline
E0/S0/Sa & $17.70 \pm 1.48$\% (120) & $13.27 \pm 1.17$\% (101)\\ \hline
Sb/Sc/Ir  & $66.37 \pm 1.81$\% (450) & $78.06 \pm 1.51$\% (594) \\ \hline
Mergers  &  $15.93 \pm 1.38$\% (108) & $8.67 \pm 1.01$\% (66) \\ \hline
\end{tabular}
\tablefoot{Uncertainties were estimated using bootstrapping with a 68\% confidence interval.}
\label{tab:gini-m20-fraction-table}
\end{table}

In the case of merger galaxies, we find that $\sim$16\% of them in the SFCGs show merger features, as expected given their dense environments. We also detected merger signatures in some FGs, likely post-mergers with disturbed morphologies, consistent with $\sim$1--5\% of nearby galaxies reported in large surveys \citep[e.g.,][]{Darg_2010,Casteels_2014,Robotham_2014}. A visual inspection of the FGs revealed mostly edge-on or diffuse systems possibly affecting the derived parameters, with no nearby companions. Conversely, merger galaxies in SFCGs display clear signs of interaction.

When comparing Tables \ref{tab:regions-fraction-table} and \ref{tab:gini-m20-fraction-table}, we found $\sim$80\% of the late-type structures (late-type and transition) in the SFCGs and the field, but only 66\% of Sb/Sc/Ir in SFCGs. While the late-type fractions are consistent when using $n$, they differ in the Gini-$M_{20}$ plane. The higher merger fraction in SFCGs nearly accounts for this discrepancy, suggesting that these groups host more LTGs undergoing mergers not evident from Sérsic indices alone.

\subsection{Physical properties for each galaxy type in the NPC}

\begin{figure}[h!]
    \centering
    \includegraphics[width=0.85\linewidth]{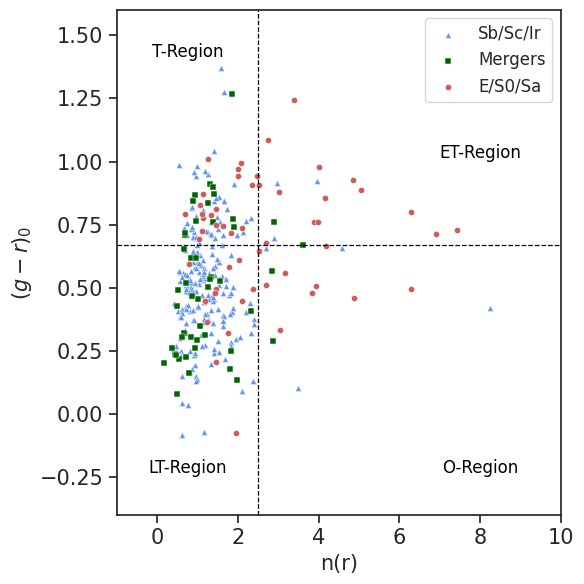}
    \caption{Star-forming compact group galaxies classified as merger (green), Sb/Sc/Ir (blue), and E/S0/Sa (red) as shown in Fig. \ref{fig:gini_m20_regions} in the Sérsic color plane.}
    \label{fig:no-param-param}
\end{figure}

In order to understand the behavior of galaxies according to their classification through non-parametric methods, we analyzed how morphological features correlate to their physical properties. Figure \ref{fig:no-param-param} shows galaxies classified in Fig. \ref{fig:gini_m20_regions} as mergers, Sb/Sc/Ir, and E/S0/Sa occupying regions in the color-\textit{n} plane. As can be seen, a majority of the red dots (E/S0/Sa) occupy the late-type, transition, and other regions, which means that there are galaxies classified as E0/S0/Sa that have blue colors and a disk shape as well as red colors and a disk shape. These galaxies should be related to early spirals (Sa) that in some cases may still be presenting star formation features. In the case of Sb/Sc/Ir, 76\% of them occupy the late-type region, while 21\% occupy the transition region. The {$\sim$} 3\% remaining are in the other and early-type regions.

The green dots in Fig. \ref{fig:no-param-param} represent galaxies with merger features in their morphology. These galaxies are displayed mainly in the late-type and transition regions, with 65\% and 27\%, respectively. Less than 8\% occupy the early-type and other regions. From their distribution in the color-\textit{n} plane, we observed that galaxies undergoing mergers in the SFCGs are mainly very blue galaxies, which means that they present a high star formation content and mainly have disk shapes. 

\begin{figure}[h!]
    \centering
    \begin{subfigure}
        \centering
        \includegraphics[width=0.95\linewidth]{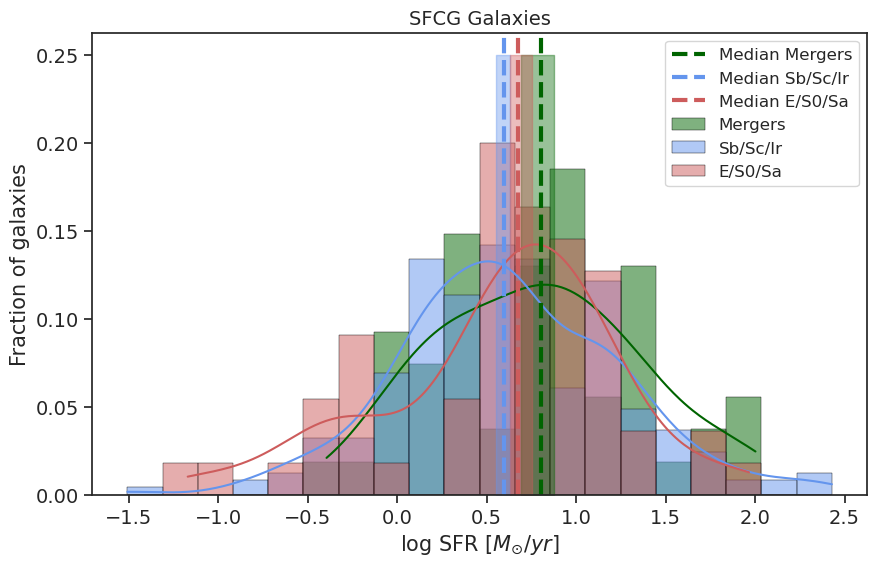}
    \end{subfigure}
    \hfill
    \begin{subfigure}
        \centering
        \includegraphics[width=0.95\linewidth]{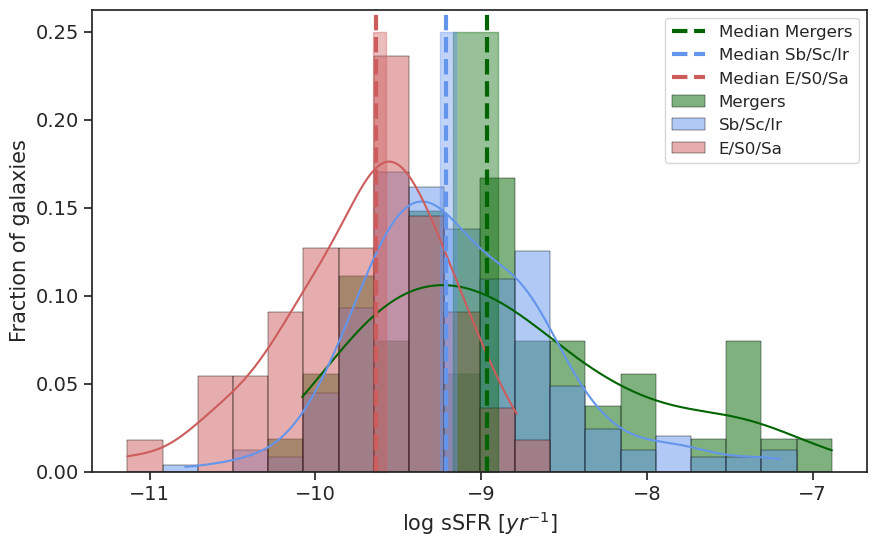}
    \end{subfigure}
    \caption{Star formation rate (top panel) and sSFR (lower panel) for galaxies in each non-parametric classification, Sb/Sc/Ir (blue), merger (green), and E/S0/Sa (red) type galaxies, in the SFCGs. Dashed lines represent median SFR values for each classification, while the shaded regions represent the errors obtained using bootstrapping to the 68\% confidence interval.}
    \label{fig:sfr_ssfr_no_param}
\end{figure}

To evaluate the evolutionary stage of galaxies, we analyzed their star formation activity based on the non-parametric classification using the same 110 SFCGs described in Sect. \ref{sec:corr_phy_prop}. Figure \ref{fig:sfr_ssfr_no_param} shows the SFR (top) and sSFR (bottom) distributions, with median values indicated by dashed lines. Merger galaxies exhibit the highest SFRs, being $0.21 \pm 0.11$ dex above Sb/Sc/Ir types, while some E/S0/Sa galaxies display blue colors and enhanced SFRs. Median log(SFR/$M_\odot$ yr$^{-1}$) values are 0.60$^{+0.03}_{-0.05}$ for Sb/Sc/Ir, 0.68$^{+0.08}_{-0.04}$ for E/S0/Sa, and 0.81$^{+0.07}_{-0.11}$ for mergers. Although the differences between Sb/Sc/Ir and E/S0/Sa are not significant, mergers tend to have higher SFRs, a result that requires a larger sample for confirmation.

The lower panel of Fig. \ref{fig:sfr_ssfr_no_param} shows the sSFR distribution for galaxies in each classification. We observed that E/S0/Sa present a significantly lower sSFR (median {$log (sSFR)= -9.63^{+0.06}_{-0.03} yr^{-1}$}) than other galaxy types, which means that the galaxies with a higher star formation content are probably more massive galaxies, so their star formation per unit mass decays. In the case of Sb/Sc/Ir, they present a higher sSFR (median {$log (sSFR)= -9.21^{+0.05}_{-0.06} yr^{-1}$}) than E/S0/Sa but lower than the merger galaxies. The merger galaxies present the highest sSFR (median {$log (sSFR)= -8.97^{+0.08}_{-0.20} yr^{-1}$}). Although, considering the uncertainties, they overlap with those in the Sb/Sc/Ir. In general, merger galaxies present a star-forming content that is higher than other galaxy types with a similar mass range to that of Sb/Sc/Ir (see the stellar mass distribution in Fig. \ref{fig:sfr-mass-no-param}). These observables suggest that merger processes enhance star formation and slightly modify the morphology of galaxies in the SFCGs.

In the case of FGs, we found that those classified as mergers do not present a higher SFR with respect to that of other galaxy types, as it happens in the SFCG sample (see Fig. \ref{fig:sfr_ssfr_no_param_cs}). It is important to consider that "merger" galaxies in the field are galaxies that present anomalies in their morphology, probably due to projection effects (edge-on galaxies), and they constitute the low-mass part of the FGs ($M_* < 10^9 M_\odot$). This is an important fact since in low-mass galaxies, the bright star-forming regions concentrate a higher fraction of the total light of the galaxy in specific regions, increasing Gini and {$M_{20}$} and driving their classification as merger galaxies.

\begin{figure}[h!]
    \centering
    \begin{subfigure}
        \centering
        \includegraphics[width=0.41\textwidth]{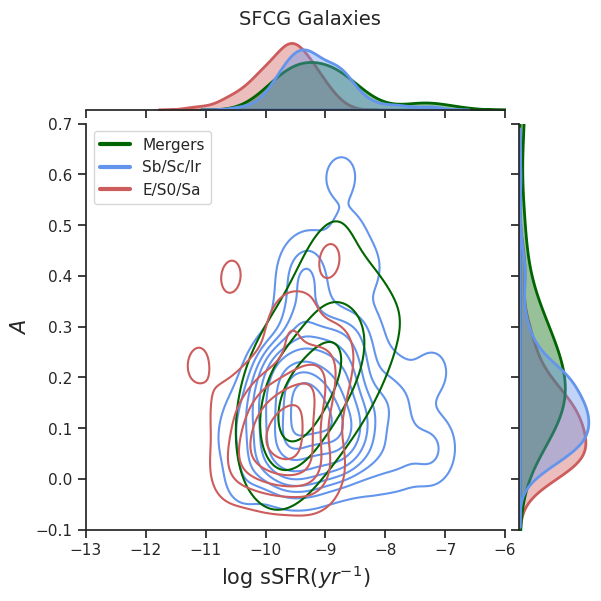}
    \end{subfigure}
    \begin{subfigure}
        \centering
        \includegraphics[width=0.41\textwidth]{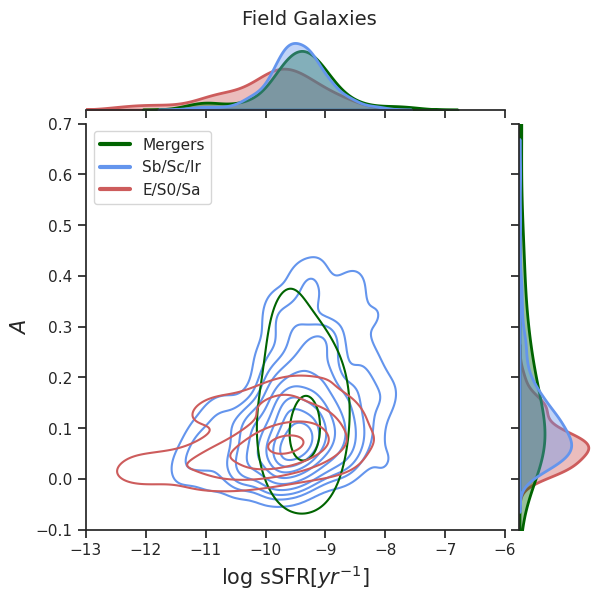}
    \end{subfigure}
    \caption{Asymmetry (A) vs. sSFR for galaxies in the SFCGs (top panel) and the field (lower panel). Lines are KDE contours for Sb/Sc/Ir (blue), merger (green), and E/S0/Sa (red) galaxies.}
    \label{fig:asym_sfcg_cs}
\end{figure}

The asymmetry (see Sect. \ref{sec:meth_non_param}) of a galaxy refers to how the light distribution of a galaxy deviates from symmetry, and it often traces disturbances caused by interactions. Figure \ref{fig:asym_sfcg_cs} presents asymmetry versus sSFR for SFCG and FGs for morphological types in the non-parametric classification. In SFCGs, asymmetry varies by type: E/S0/Sa have the lowest median asymmetry ($0.11^{+0.01}_{-0.01}$), followed by Sb/Sc/Ir ($0.13^{+0.01}_{-0.01}$) and mergers ($0.21^{+0.03}_{-0.02}$). Perturbed galaxies show elevated asymmetry values due to tidal features from interactions in dense environments such as CGs. These features can be faint, highlighting the importance of deep imaging for their detection.

In the case of galaxies in the field (bottom panel), through marginal plots we observed that asymmetry is not significantly different between the merger and Sb/Sc/Ir galaxies, while we observed that E/S0/Sa galaxies do actually present lower asymmetry values with respect to those of the other populations. This suggests that FGs classified as mergers may not actually have merger features and the high Gini and $M_{20}$ values could therefore be associated with projection effects, or these galaxies are post-merger galaxies that have undergone interactions not visible in this stage. We show the asymmetry distribution of galaxies for each non-parametric classification in Fig. \ref{fig:asym_distribution}.

\section{Discussion}\label{sec:discussion}

\subsection{Implications of the environment on the morphological properties of galaxies}

One of the earliest isophotal analyses of CG galaxies was conducted by \cite{MdeOliv_1994} on 202 galaxies in 92 Hickson compact groups (HCGs; \citealt{Hickson_1982}). They revealed larger ellipticals, distortions, and interaction features in 43\% of the galaxies compared to those in less dense environments. Later, \cite{Coenda_2012} found that CG galaxies are slightly more compact than FGs (lower $R_e$). These studies focused on dynamically evolved Hickson-like CGs. In contrast, our SFCG sample is dominated by disk-like systems, with $\sim$88\% of galaxies having $n < 2.5$. Consistent with \cite{MdeOliv_1994}, we find ETGs in SFCGs to be slightly larger than their field counterparts, as given by the median values of the $R_e$ in Table \ref{tab:param_values_regions}. Nevertheless, this is based on a small ETG subsample ($\sim$ 6\%) and large uncertainties ($\sim$ 20\%). Conversely, LTGs in SFCGs show median $R_e$ values $\sim$0.5 kpc larger than those in the field. Although assigning group redshifts to galaxies without spectra may introduce small biases, the median $R_e$ values remain consistent with previous studies, excluding a few outliers likely affected by measurement errors.

\cite{Coziol_2007} analyzed 25 galaxies in eight HCGs, finding that 52\% showed merger signatures and morphological asymmetries due to inhomogeneous stellar mass distributions, likely from interactions, mostly "dry" mergers with little gas and no significant star formation. In comparison, only 16\% of galaxies in our SFCGs show merger features, a lower fraction than reported in previous studies \citep{MdeOliv_1994, Coziol_2007}. Nonetheless, we also observed a higher asymmetry in SFCG galaxies relative to FGs, consistent which is with the results of \citeauthor{Coziol_2007}. This supports the idea that interaction-driven morphological disturbances, particularly in merger candidates, contribute to increased asymmetry in compact group environments.

\cite{Montaguth_2023} conducted a similar analysis on more dynamically evolved CGs, examining the $n$ and {$R_e$} parameters in the the color–$n$ plane. They found a clear bimodality in the transition region (particularly in $n$) that was absent in FGs, suggesting environment-driven morphological changes. In our study, no such bimodality is observed, though the transition galaxies show a fraction with $n > 1.5$, which is higher compared to that in LTGs but is not significantly different compared to the FGs. Additionally, \cite{Montaguth_2024} reported that galaxies in isolated CGs are more concentrated in the $n$–$R_e$ space for $n < 1.75$. This suggests that galaxies in SFCGs may be undergoing mild morphological transformations, resembling the behavior seen in isolated CGs rather than in more evolved systems.  

\subsection{Implications of the environment on the physical properties of galaxies}

Several studies have examined environmental effects on galaxies in CGs from different perspectives. \citet{Walker_2010, Walker_2012, Walker_2013} reported a gap in the mid-infrared 3.6–8\,$\mu$m IRAC color, indicating a deficit of mid-infrared transition galaxies (MIRTGs) and suggesting environment-driven star formation suppression. \citet{Alatalo_2015} found similar evidence from CO maps of 12 HCGs, where shocks inject turbulence into the molecular gas, preventing its collapse and thus quenching star formation without gas expulsion. Using a larger sample of 1770 Hickson-like CGs, \citet{Bitsakis_2016} showed that star formation rates decreased by a factor of three to ten over the past 3\,Gyr and that CG galaxies have distinct star formation histories compared to the field, reinforcing the environmental impact. Apart from gas stripping, turbulence and shocks appear crucial in this suppression. In our analysis of SFCG and FGs, we find no quenched members in SFCGs following \citet{Wetzel_2013}, unlike FGs and the more dynamically evolved CGs studied by \citet{Montaguth_2023, Montaguth_2024}, where $\sim$10\% of LTGs and over 50\% of all CG galaxies are quenched.

We found that 65\% of merger galaxies occupy the late-type region, and 27\% of them occupy the transition region. Analyzing the behavior of the SFR of galaxies according to the non-parametric classification, we observed that merger galaxies present the highest median SFR and sSFR (see Fig. \ref{fig:sfr_ssfr_no_param}); however, SFR is similar to Sb/Sc/Ir, with similar mass ranges (see Fig. \ref{fig:sfr-mass-no-param}). Contrary to what has been observed in different studies from \cite{Walker_2010, Alatalo_2015}, and \cite{Bitsakis_2016} in more evolved CGs, we claim that merger processes in the SFCGs are not (yet) suppressing the star formation of galaxies but instead are enhancing it slightly, with merger galaxies producing an average of $\sim1.5$ solar masses per year more than Sb/Sc/Ir with a similar galaxy stellar mass range. 

The CGs play a key role in galaxy evolution and large-scale structure assembly. \citet{Cortese_2006} studied a BIG in a local cluster, finding disturbed morphologies and enhanced H$\alpha$ emission, which is indicative of starburst activity triggered by both local and global environments. Furthermore, they suggested that pre-processing significantly contributes during the early stages of cluster assembly. In this sense, our results agree with those reported by these authors and support the scenario in which the local environment of galaxies is responsible for physical and morphological transformations of galaxies.

\cite{Montaguth_2024} analyzed the star formation activity of galaxies in CGs as a function of their larger-scale environment, finding that ETGs and transition galaxies in non-isolated CGs exhibit lower sSFRs and a higher early-type fraction compared to those in isolated CGs. They also reported that LTGs evolve differently depending on the environment, with star formation quenching being more efficient in non-isolated CGs—likely due to additional mechanisms such as RPS. Although our SFR estimates were obtained through a different methodology, we observed a consistent trend: The sSFR distributions of all morphological types in our SFCG sample are $\sim$32–37$\%$ higher than those of their field counterparts, suggesting that SFCGs are at an earlier evolutionary stage with respect to all previous samples where star formation is still active.

\cite{Lopes_2024} analyzed pre-processing in 153 local clusters ($z \leq 0.1$) by comparing star-forming and late-type galaxy fractions across various environments. They found that the star-forming fraction ($f_{SF}$) is at least 14\% lower in substructures and isolated groups at $r \geq 3-5r_{200}$ compared to the field, indicating pre-processing. The effect increases with substructure richness but remains significant even for pairs and triplets. They also reported that the late-type fraction ($f_{LT}$) is consistently lower in dense regions and that morphological transformation occurs on longer timescales than quenching. While our findings differ, they could be reconciled since \cite{Lopes_2024} showed that smaller groups are more star-forming, and they did not study compact groups, where different physical processes may be at play. Also, since we do not observe significant morphological changes in the SFCGs, our results agree that the morphological transformations occur more slowly than the effects on the star formation, as mentioned at the end of Sect. \ref{sec:morph_properties}.  

We observed that SFCGs behave differently from other CGs according to their physical properties, with a star-forming activity higher than that in the field and merger processes enhancing the star formation instead of suppressing it. It is possible that in the SFCGs we are observing a different stage in the evolution of these structures, i.e., a snapshot in a different moment in the lives of CGs. This will be further explained in Sect. \ref{sec:evolutionary_stage}.

\subsection{SFCGs as an evolutionary stage of compact groups of galaxies}\label{sec:evolutionary_stage}

\cite{Hernandez-Fernandez_2015} defined SFCGs as structures with a velocity dispersion lower ($\sigma_{l-o-s} \sim 120 km/s$) than other CG samples, a small crossing time ($H_{0}t \sim 0.05$), and high star formation content. The authors also found that 26 out of 280 ($\sim 9\%$) SFCGs are located in galaxy cluster infall regions (although they do not mention which groups they are). Here, we were able to quantify the SFR, stellar mass, sSFR, and morphological features in a parametric and non-parametric approach and to classify galaxies according to these properties. This new information allowed us to characterize the behavior of galaxies in the SFCGs and the effects of the environment in the context of galaxy evolution in CGs.

\cite{Montaguth_2024} have built an evolutionary scenario for CGs. They propose that the major structures where CGs are embedded enhance the quenching processes and morphological transformations of galaxies in these substructures so that earlier stages of the evolution present a higher late-type fraction. They also find that physical and morphological differences exist between galaxies in isolated and non-isolated CGs since the sSFR of ETGs and transition galaxies in isolated CGs is higher, and transition galaxies in CGs populate the $n < 1.75$ region more densely in the $n-R_e$ plane. Galaxies in non-isolated CGs contribute to the higher $n$ part of the bimodality. Their approach to a major structure also influences the dynamics of the CGs since non-isolated CGs present a higher velocity dispersion ($\sigma < 800 km/s$) than that of isolated CGs ($\sigma \leq 400 km/s$), suggesting that the dynamics of non-isolated groups is influenced by the dynamics of the major structure they belong to. \cite{Hagar_2023} observed something similar using simulations and found that the dispersion of groups increases as they approach galaxy clusters, and groups become more compact as they fall in. 

According to \cite{Hernandez-Fernandez_2015}, SFCGs exhibit a high fraction of star-forming galaxies, have low velocity dispersions, and present moderate crossing times, indicating a less dynamically evolved stage compared to other CGs. Galaxies in these groups form stars at rates higher than those of FGs, with mergers further enhancing their SFR. Morphologically, they resemble transition galaxies in isolated CGs, with a concentration at low $n_r$ and no evidence of bimodality or strong morphological transformations. These findings suggest that SFCGs represent an early evolutionary phase in CGs, where star formation has not yet been quenched. Most of them are not significantly influenced by the dynamics of larger structures, although a small fraction may lie near cluster infall regions, implying that the observed SFR excess is driven mainly by internal gravitational interactions. In this context, the SFCGs are dynamically less evolved than other CG samples.

\section{Conclusions}\label{sec:conclusion}

In this paper, we have studied a sample of SFCGs to investigate how the environment affects galaxy evolution through their physical and morphological properties. We performed a morphological analysis with \texttt{GALFITM} to obtain the Sérsic index (\textit{n}) and effective radius ($R_e$) for 640 galaxies in 207 SFCGs with DECaLS data, out of the 280 identified by \cite{Hernandez-Fernandez_2015}. We derived physical properties (SFR and stellar mass) for 356 galaxies in 110 SFCGs with spectroscopic data, representing half of the sample. Galaxies were classified by type following \cite{Vika_2015}, adopting $(g - r) = 0.67$ as the color threshold due to the lack of \textit{u} band imaging in DECaLS. All subsequent analyses were based on the 110 SFCGs with available spectroscopic redshifts for at least one member, and we applied K-corrections to ensure consistency in the derived magnitudes.

In addition, we studied the morphology of the whole available sample with a non-parametric approach to obtain parameters such as Gini, $M_{20}$, and asymmetry. We separated the galaxies by their morphology only, especially the galaxies showing merger characteristics. We were able to correctly extract the parameters for 678 galaxies belonging to SFCGs, finding that 16\% of galaxies are classified as mergers. We then studied the physical properties of galaxies in this classification for the same 110 SFCGs mentioned above.

The morphology and other physical properties obtained for the galaxies in the SFCGs were also derived for a control sample composed of 743 FGs that have similar stellar masses and redshift distributions as those of the SFCG sample. In addition, we applied the same color cut used to classify the SFCGs in \cite{Hernandez-Fernandez_2015}, in order to avoid biases and study the environmental effect.

We summarize our main results as follows:

\begin{enumerate}
    \item We find no significant morphological differences between galaxies in SFCGs and the field. The Sérsic index shows no bimodality in the ``transition region,” unlike more evolved CG samples \citep{Montaguth_2023, Montaguth_2024}, although a similar behavior to transition galaxies in isolated CGs is observed in SFCGs with galaxies more concentrated at $n_r < 1.75$. With respect to effective radius, LTGs in SFCGs are statistically larger than those in the field, contrary to \cite{Coenda_2012}. 
    \item Considering the non-parametric approach, we find that 16\% of galaxies in the SFCGs show merger features, with a consistently higher asymmetry value. Although we find 8\% of "merger" galaxies in the FGs, visual inspection indicates that these galaxies are effectively isolated, and these cases are related to projection effects and low-mass irregular galaxies. These FGs also have low values of asymmetry, which reinforces the idea of a misclassification.
    \item Merger galaxies in the SFCGs show an enhancement in SFR compared to E/S0/Sa and Sb/Sc/Ir galaxies, with both SFR and sSFR values being higher.
    
    \item Considering the group characteristics and the galaxies within them, we speculate that SFCGs represent an earlier evolutionary stage in the CG sequence proposed by \cite{Montaguth_2024}. In this stage, CGs are less dynamically evolved, with lower velocity dispersions and moderate crossing times, compared to other samples. At this point, galaxy mergers enhance rather than suppress star formation, and the morphological behavior resembles that found in isolated CGs. This transforms SFCGs into a good place to study the initial effects of interactions in dense environments. 
\end{enumerate}

To better understand the mechanisms regulating star formation in SFCGs, we aim to perform a detailed spectroscopic follow-up using Gemini/GMOS to analyze the chemical composition and ionized gas. Future H\textsc{I} observations will also be essential to study the intragroup medium and its impact on star formation. Also, confirming group membership spectroscopically is necessary to rule out interlopers and confirm membership of other non-star-forming galaxies that may be a part of the SFCGs and are not being considered as a result of the selection method. Finally, we plan to expand the SFCG sample, emphasizing their role in the evolution of galaxies in compact environments.

The connection between SFCGs and larger structures may influence galaxy evolution and shed light on the roles of mergers and RPS. Pre-processing is often studied by comparing star-forming fractions in clusters and the field and by identifying substructures in cluster outskirts. However, obtaining a full understanding requires combining cluster and group analyses, as evolutionary processes may differ. Studying early-stage systems such as SFCGs is therefore essential to constraining pre-processing scenarios and the physical mechanisms driving galaxy evolution. 

\begin{acknowledgements}
S.O-G., A.M. and R.D. gratefully acknowledge support from the ANID BASAL project FB210003. S.T-F. acknowledges the financial support of ULS/DIDULS through a regular project number PR2453858. A.M. acknowledges support from the ANID FONDECYT Regular grant 1251882 and funding from the HORIZON-MSCA-2021-SE-01 Research and Innovation Programme under the Marie Sklodowska-Curie grant agreement number 101086388. We acknowledge Alvarez-Candal A. for his comments on this work. We gratefully acknowledge the anonymous referee for insightful comments that helped improving this paper.
\end{acknowledgements}

\bibliographystyle{aa}
\bibliography{bibliography}

\begin{appendix}

\FloatBarrier
\section{GALFITM modeling and morphological properties}\label{sec:appendix_galfitm}

``MorphoPLUS" is designed to be used over S-PLUS data, from the image download to the GALFITM software execution. Since we used DECaLS data, we added some extra steps. We downloaded DECaLS science and weight images as it was explained in the previous section, then we ran SExtractor over detection images built by a composition of the \textit{g, r, z} bands, extracting image and WCS coordinates to crossmatch with the galaxies catalog, and different properties that are useful in the configuration file of GALFITM (\texttt{FLUX\_RADIUS, KRON\_RADIUS, ELONGATION} and \texttt{THETA\_IMAGE}). From SExtractor we obtained a segmentation map, masking all the sources detected including galaxies we are interested in. Then, the code utilizes those segmentation maps to produce new ones by unmasking those objects of interest, assigning a numerical value of 1 to all those pixels that are masked, and 0 to all pixels that are not. This is because GALFITM requires  a mask containing values of 0 to all the regions to be fitted. Finally, the PSF was computed utilizing the \texttt{Photutils} python package, in which we selected stars in a 25 pixels box constraining for magnitude and a small half-light radius. After filtering for stars with no other source of contamination nearby, we also constrain for S/N in the lower and upper limits, to avoid stars with a poor S/N and saturated stars. Then, photutils builds an effective PSF with those selected stars. We repeated this process for each SFCG.

\begin{table}[h!]
\caption{Median of the morphological parameters for each region defined in Sect. \ref{sec:gal_classification}.}
\centering
\resizebox{0.8\linewidth}{!}{ % escala al 90% del ancho del texto}
\begin{tabular}{|c|c|}
\hline
\textbf{Galaxy Type} & \textbf{Median \textit{n}} \\ \hline
Early Type    & $3.63 \pm 0.29$ ($4.23 \pm 0.01$) \\ \hline
Late Type     & $1.09 \pm 0.05$ ($1.10 \pm 0.02$) \\ \hline
Transition    & $1.29 \pm 0.06$ ($1.22 \pm 0.03$) \\ \hline
Other         & $3.89 \pm 0.40$ ($3.58 \pm 0.01$) \\ \hline
Whole Sample  & $1.23 \pm 0.04$ ($1.26 \pm 0.02$) \\ \hline

\textbf{Galaxy Type} & \textbf{Median $R_e$ (kpc)} \\ \hline
Early Type    & $7.67 \pm 1.38$ ($5.42 \pm 0.40$) \\ \hline
Late Type     & $4.18 \pm 0.15$ ($3.78 \pm 0.17$) \\ \hline
Transition    & $5.42 \pm 0.35$ ($5.09 \pm 0.28$) \\ \hline
Other         & $2.24 \pm 0.76$ ($3.08 \pm 0.01$) \\ \hline
Whole Sample  & $4.49 \pm 0.14$ ($4.26 \pm 0.01$) \\ \hline
\end{tabular}}
\tablefoot{Values correspond to the SFCG galaxies, and in parenthesis we observe values for FGs. Uncertatinties are estimated using bootstrapping with a 68\% confidence interval.}
\label{tab:param_values_regions}
\end{table}

\begin{figure*}[h!]
    \centering
    \includegraphics[width=1\linewidth]{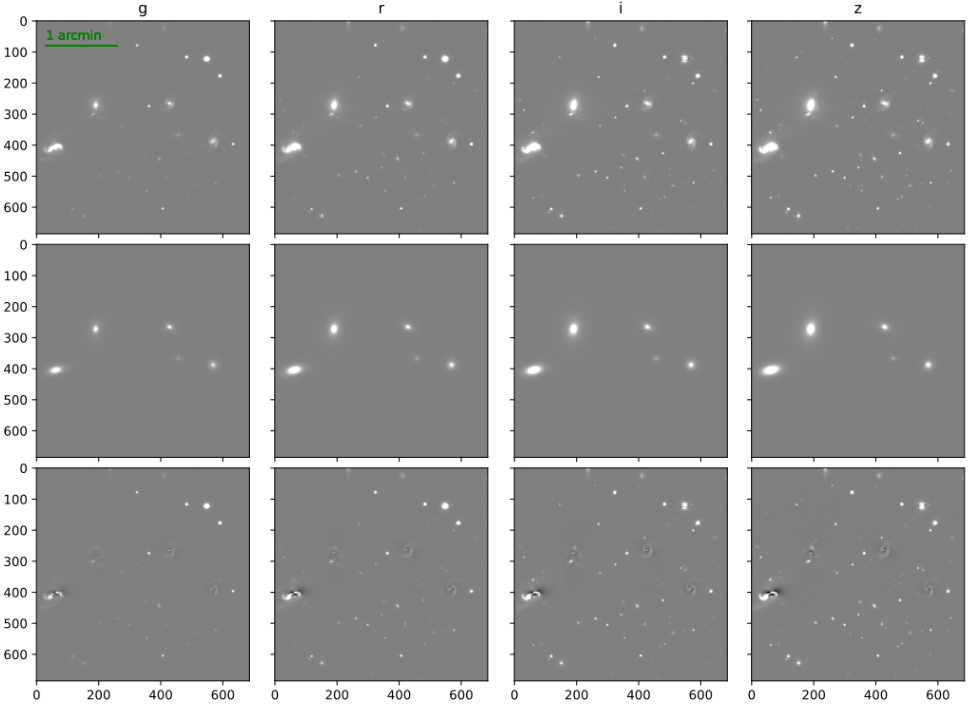}
    \caption{Example of GALFITM modeling over SFCG 235. Upper row is for the input images in each filter, middle row is the model image from GALFITM and lower row is the substraction between the input image and the model (residual image).}
    \label{fig:galfitm}
\end{figure*}

Subsequently, the code runs GALFITM using the configuration files built before. In this case we used the GALFIT 1.4.4 version, which can be downloaded in the MegaMorph official website\footnote{\url{https://www.nottingham.ac.uk/astronomy/megamorph/}}. The GALFITM output we obtained consists in the input image, the model and the residual (substraction between the image and the model) for each photometric band (see Fig. \ref{fig:galfitm}). Then the code reads the header of the model files, extracting and tabulating the Sérsic index value, effective radius, position angle, magnitude, position and axis ratio, with their respective uncertainties. Together with this, a .svg image is produced for each group, containing the image, model and residual in each filter.

\begin{figure}[h!]
    \centering
    \begin{subfigure}
        \centering
        \includegraphics[width=0.3\textwidth]{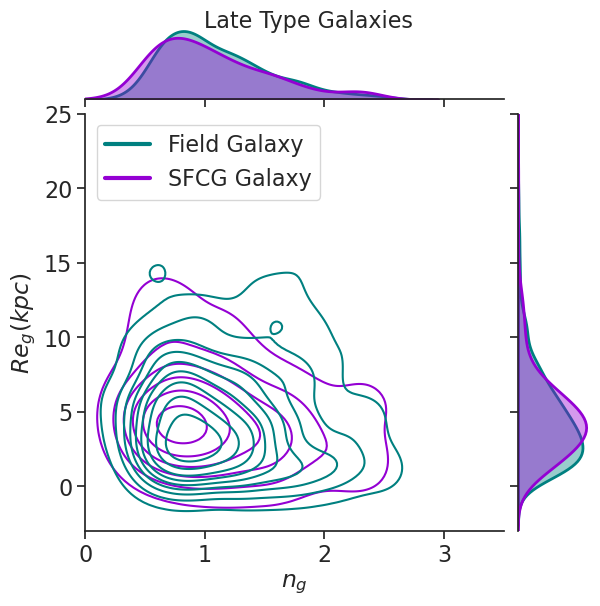}
    \end{subfigure}
    \begin{subfigure}
        \centering
        \includegraphics[width=0.3\textwidth]{images/late-n-re-r.png}
    \end{subfigure}
    \begin{subfigure}
        \centering
        \includegraphics[width=0.3\textwidth]{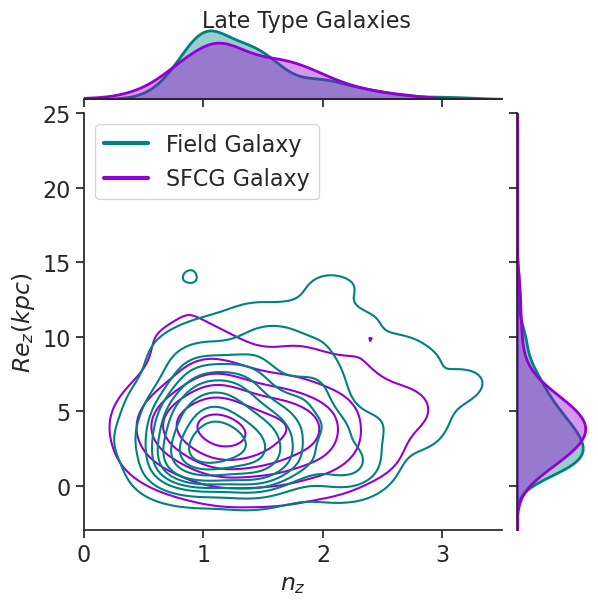}
    \end{subfigure}
    \caption{Effective radius vs. Sérsic in the \textit{g}, \textit{r}, and \textit{z} filters for the LTGs in the SFCGs (violet) and the field (teal).}
    \label{fig:late_n_re_all}
\end{figure}

\begin{figure*}[h!]
    \centering
    \begin{subfigure}
        \centering        \includegraphics[width=0.3\textwidth]{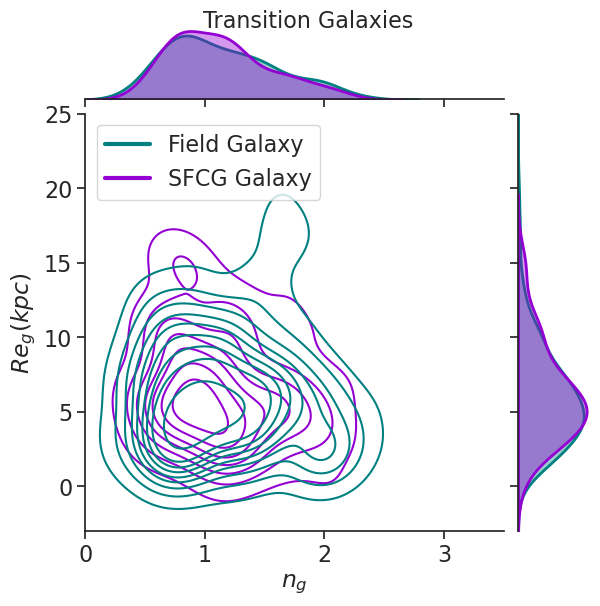}
    \end{subfigure}
    \begin{subfigure}
        \centering        \includegraphics[width=0.3\textwidth]{images/transition-n-re-r.png}
    \end{subfigure}
    \begin{subfigure}
        \centering        \includegraphics[width=0.3\textwidth]{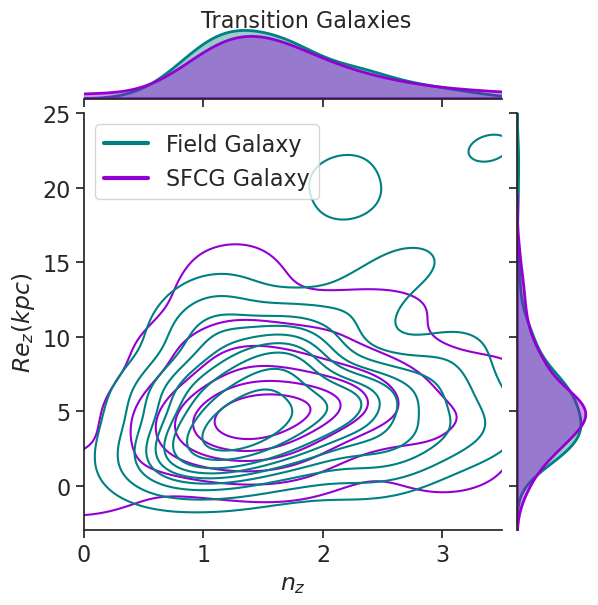}
    \end{subfigure}
    \caption{Effective Radius vs. Sérsic in the \textit{g}, \textit{r}, and \textit{z} filters; for the transition galaxies in the SFCGs (violet) and the field (teal).}
    \label{fig:trans_n_re_all}
\end{figure*}

We have observed the behavior of the morphological parameters in the g, r and z bands for LTGs and transition galaxies (Figs. \ref{fig:late_n_re_all} and \ref{fig:trans_n_re_all}, respectively). We do not witness significant changes in the $R_e$ to redder bands for any of the samples, although we observe that $n$ increases mainly in the \textit{z} band. The latter is expected, since in redder wavelengths we are observing mainly the stellar content of the galaxy rather than the higher energy photons coming from massive stars in the star-forming regions in the galactic disks, so the light in that band would be more concentrated and we will be obtaining a higher $n$.

\begin{figure*}[h!]
    \centering
    \includegraphics[width=0.85\linewidth]{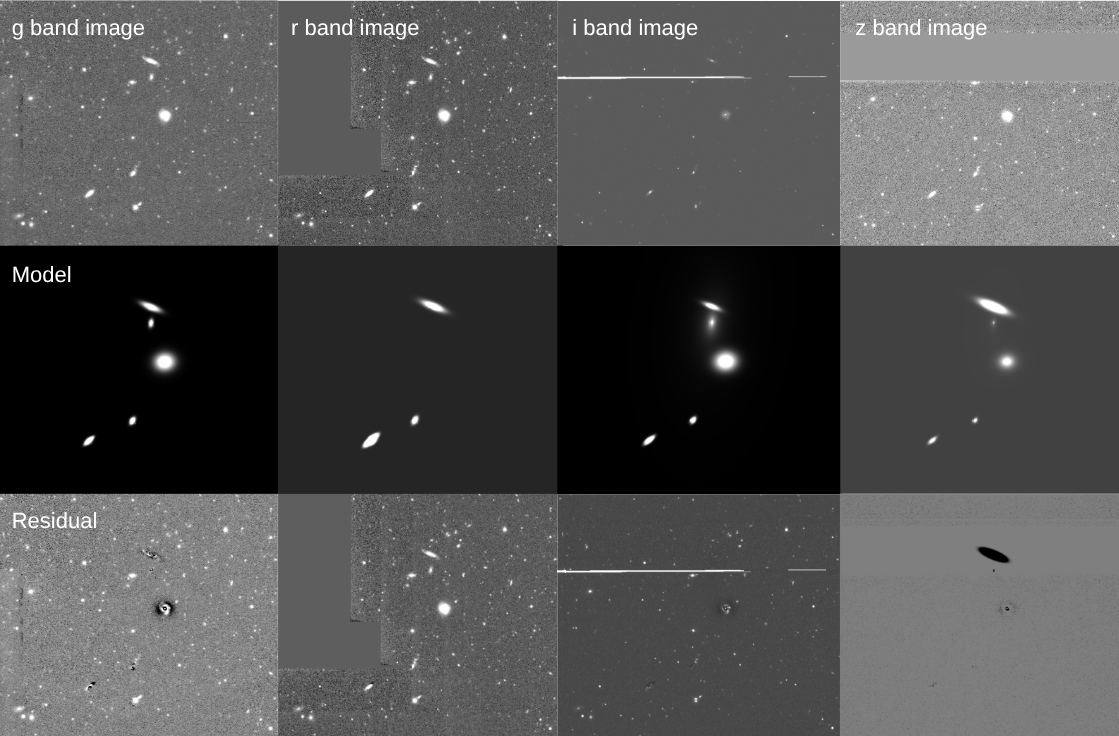}
    \caption{Example of a failed GALFITM modeling over SFCG 228, due to an incomplete input image. Upper row is for the input images in each filter, middle row is the model image from GALFITM and lower row is the substraction between the input image and the model (residual image).}
    \label{fig:bad_fit}
\end{figure*}
\newpage
%\FloatBarrier
\section{Correlation between methods of obtaining physical properties
}\label{sec:appendix_phy_properties}

In this section we show the correlation between the photometric methods to obtain physical properties of galaxies and the information available from SED-fitting derived properties from the GWSLC.

\begin{figure}[h!]
    \centering
    \includegraphics[width=0.85\linewidth]{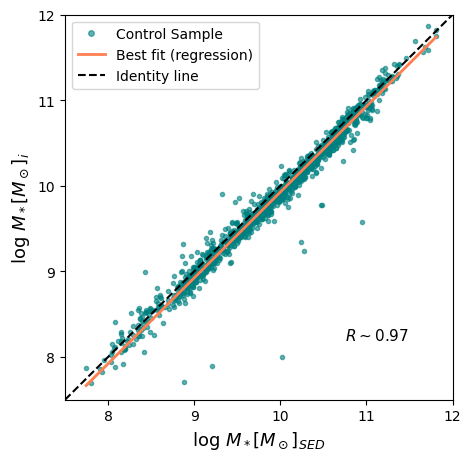}
    \caption{Stellar mass obtained using (\textit{g - i}) color and $M_{i}$ vs. stellar mass computed with SED fitting from the GWSLC. The orange line represents the best fit from the linear regression with a correlation coefficient of R $\sim$ 0.97. The dashed black line represents the Identity line.}
    \label{fig:photvssed}
\end{figure}

\begin{figure}[h!]
    \centering
    \includegraphics[width=0.85\linewidth]{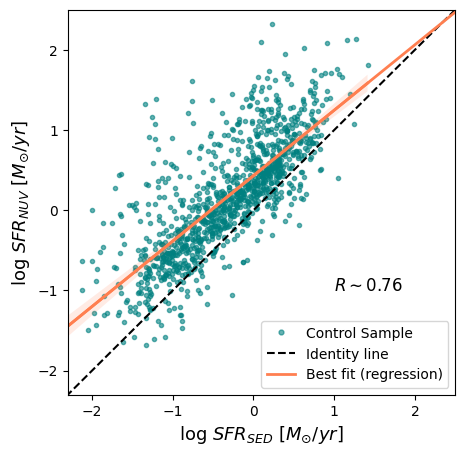}
    \caption{Star formation rate obtained using $L_{NUV}$ vs. SFR obtained through SED fitting from the GWSLC. The orange line represents the best fit from the linear regression with a correlation coefficient of R $\sim$ 0.76. The dashed black line represent the Identity line.}
    \label{fig:sfr-correlation}
\end{figure}

\newpage
\section{Color limit using (\textit{g - r}) for galaxy classification}\label{sec:appendix_classification}

\begin{figure}[h!]
    \centering
    \includegraphics[width=1\linewidth]{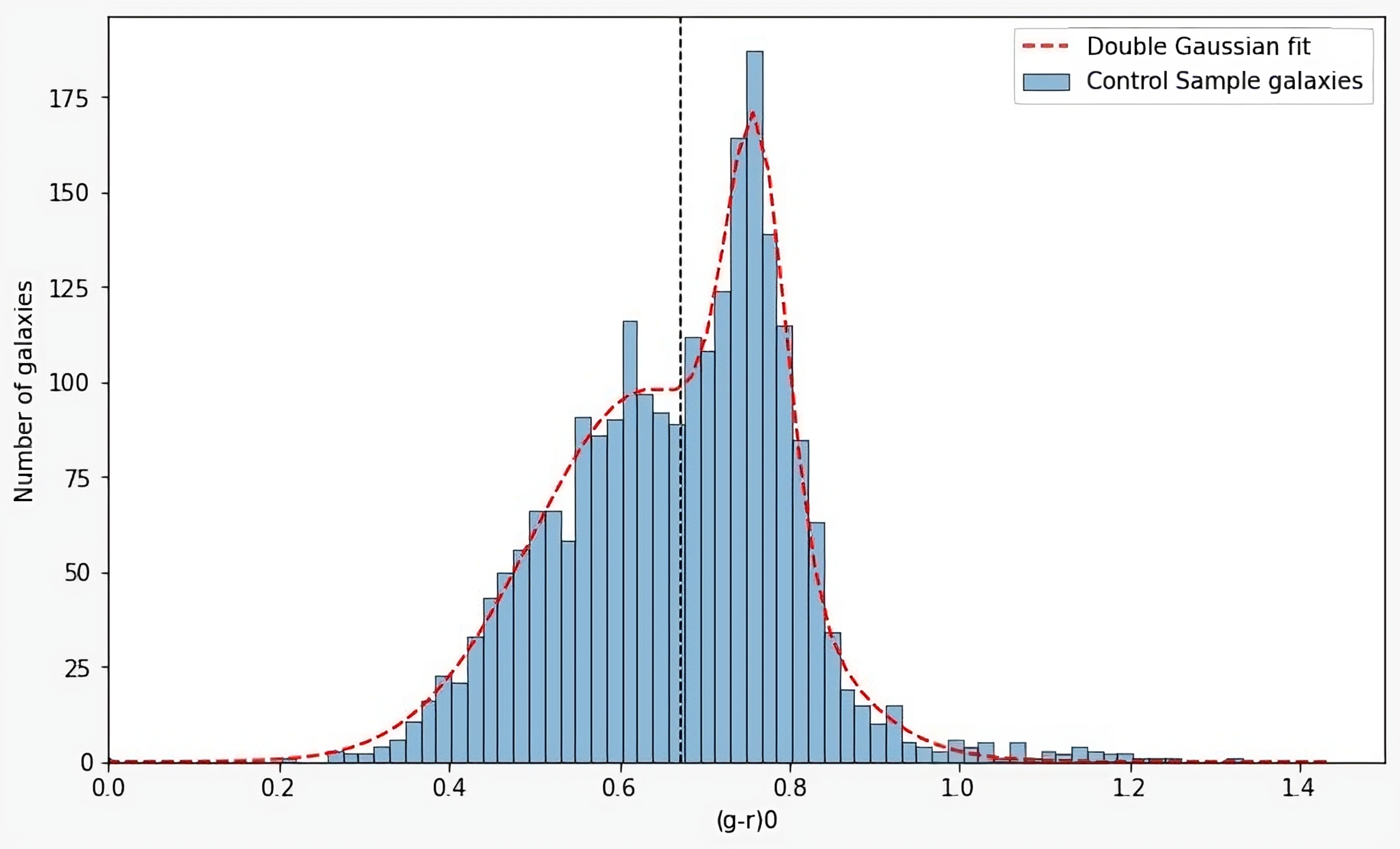}
    \caption{Double Gaussian fitting over the color distribution of the FGs. The vertical dashed line represents the two Gaussians' interception at rest-frame color (\textit{g - r}) = 0.67. Figure credit: Gissel P. Montaguth.}
    \label{fig:doublegaussian}
\end{figure}

DECaLS provide images in the \textit{g, r, i, z} filters, which is why we cannot use exactly the same color definition as in \cite{Vika_2015}. Analogously, in this work we used the (\textit{g - r}) color, considering the color limit as (\textit{g - r}) = 0.67. This value was obtained by a double Gaussian fitting over the color bimodality of a sample of FGs in the STRIPE-82 S-PLUS field, as shown in Fig. \ref{fig:doublegaussian}, setting the limit as the value in which both gaussians intercept.  

\begin{figure}[h!]
    \centering
    \includegraphics[width=0.9\linewidth]{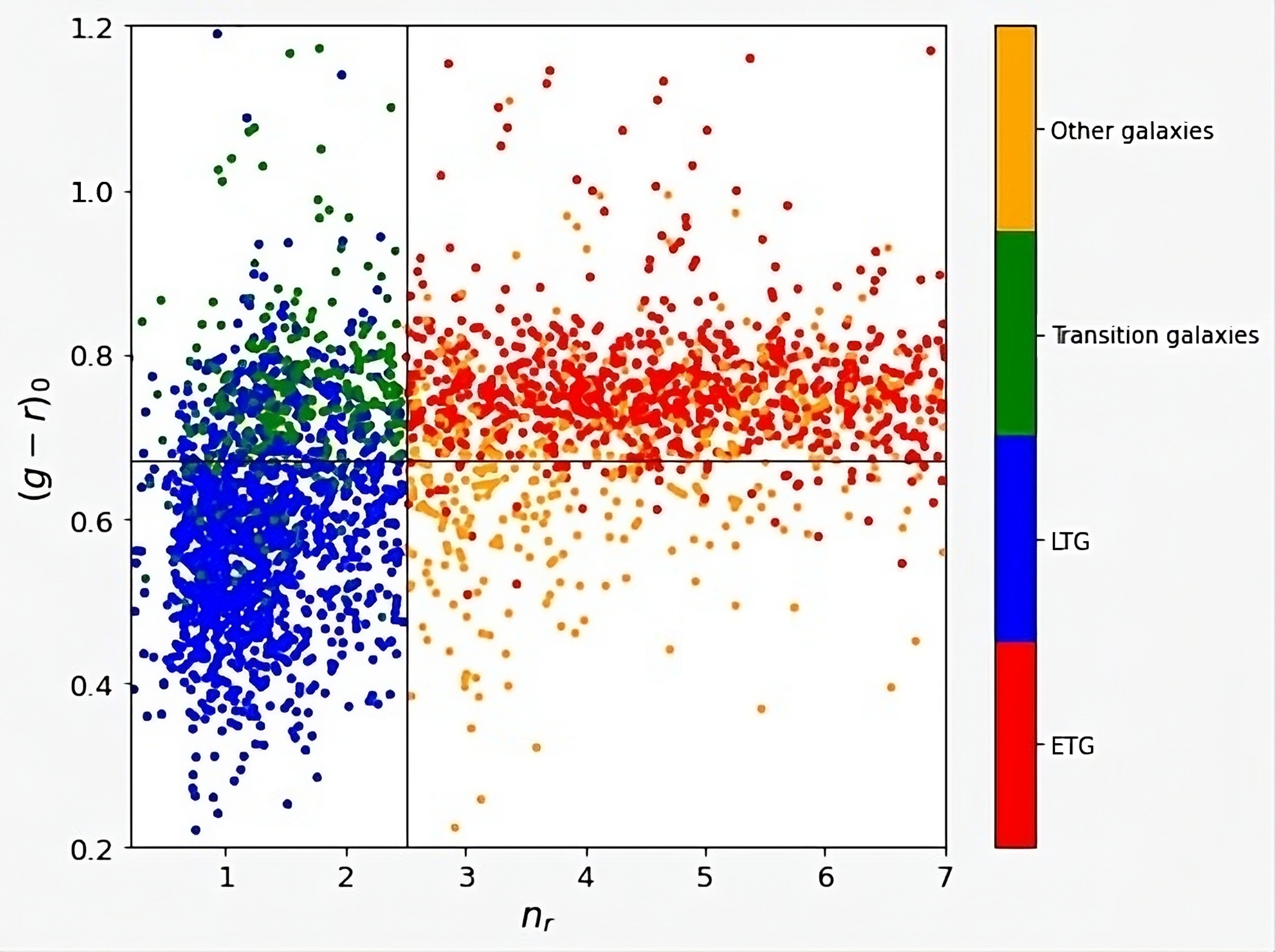}
    \caption{Different galaxy types previously classified using \cite{Vika_2015} limits, located over the (\textit{g - r})-\textit{n} plane. Vertical line is \textit{n} = 2.5, and horizontal line (\textit{g - r}) = 0.67. Figure credit: Gissel P. Montaguth.}
    \label{fig:grdefinition}
\end{figure}

It is worth noting that using this value include a little contamination over some of the regions, as seen in Fig. \ref{fig:grdefinition}. Here we show the same galaxies than in Fig. \ref{fig:doublegaussian}, classified as each type using the (\textit{u - r}) color limit, plotted over the $n_r$ versus (\textit{g - r}) plane. The region where we find the more significant overlapping is in the upper right part of the plane, in the ETGs region, where we can see a mixing between ETGs and other galaxies. Given the nature of the galaxies in the SFCGs, we do not expect to find many galaxies in those regions, since most of the galaxies should be LTGs.

\section{Morphological parameters in DECaLS and S-PLUS}
\label{sec:app_morph-param-spdec}

The structural parameters of galaxies can be strongly affected by the quality of the images and the filters used to obtain them. In this section we aim to analyze the behavior of morphological properties of galaxies in S-PLUS and DECaLS surveys. The S-PLUS survey contains 12 filters, which theoretically would be an advantage in GALFITM modeling, hoping that with a major amount of filters we will constraint the parameters with more accuracy. In the other hand, DECaLS only contains four broadband filters, but it is deeper and has better resolved images.

In order to compare the performance of GALFITM over each survey, we computed the median of the morphological parameters in each broadband filter to see if there are any differences between them. It is important to mention that uncertainties given by GALFITM are one magnitude larger for S-PLUS data than for DECaLS, which is probable due to the lower S/N in S-PLUS compared to the DECaLS images. However, we believe that the GALFITM uncertainties are underestimated, and adding statistical uncertainties to the data provides reliability to the results.

\begin{figure}[h!]
    \centering
    \includegraphics[width=\linewidth]{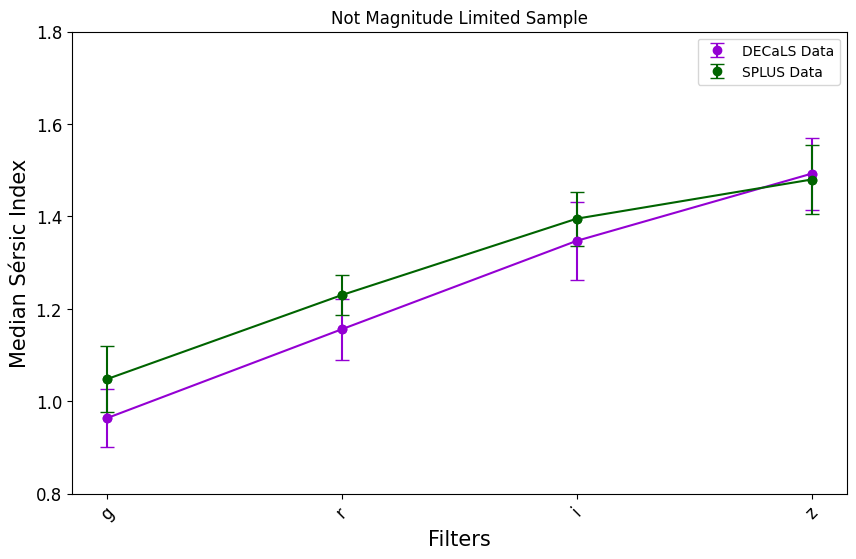}
    \caption{Median Sérsic index for galaxies in DECaLS (violet) and S-PLUS (green) surveys, not limiting the sample by magnitude, error bars are obtained using the bootstrapping method, with 1000 re-samples.}
    \label{fig:nnolimit}
\end{figure}

\begin{figure}[h!]
    \centering
    \includegraphics[width=\linewidth]{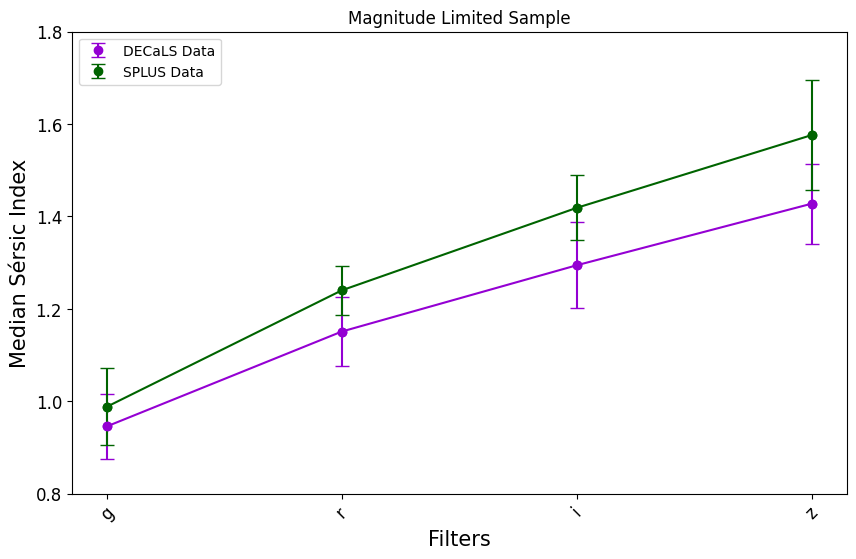}
    \caption{Median Sérsic index for galaxies in DECaLS (violet) and S-PLUS (green) surveys, limiting magnitudes up to 18.5, error bars are obtained using the bootstrapping method, with 1000 re-samples.}
    \label{fig:nlimit}
\end{figure}

In Fig. \ref{fig:nnolimit} we can see the behavior of \textit{n} where there is not limit in magnitudes. We find that the median Sérsic index from \textit{g} to \textit{z} filter increases its value in 0.53 and 0.44 for DECaLS and S-PLUS, respectively. This is expected, given the fact that to redder wavelengths we are expecting to see the main and old stellar component of a galaxy, which is more concentrated in the bulge than in the disk of a LTG (which we should find in the SFCGs), so the \textit{n} should increase. Practically the same difference is seen when we only observe galaxies brighter than \textit{r} = 18.5 (magnitude limited sample, as shown in Fig. \ref{fig:nlimit}), once we analyze the change of the median \textit{n} with filters. With respect to the difference between each survey, in both samples (limited and not limited by magnitude) we can see that \textit{n} in S-PLUS is systematically higher, particularly in the magnitude limited sample in which is higher in all filters. On average, the difference between both surveys is \textit{n} = 0.11 {$\pm$} 0.01 for the magnitude limited sample.  

Regarding the effective radius, we find that the trend is consistent in the sample limited by magnitude and not limited by magnitude for the DECaLS data, in which {$R_e$} monotonically decreases to the redder filters. In the case of S-PLUS, in the not magnitude limited sample it decreases from the \textit{g} to \textit{r} filter and it slighlty increases to redder filters. Although, for the S-PLUS data the trend is flatter, since the difference in the {$R_e$} is 0.12 $\pm$ 0.02 and 0.19 $\pm$ 0.02 for the not magnitude limited and magnitude limited sample, respectively. In the DECaLS data, the difference from the \textit{g} to \textit{z} filter is 0.41 $\pm$ 0.05 and 0.35 $\pm$ 0.07 for the not magnitude limited and the limited sample, respectively. The decrement of the {$R_e$} to redder colors is expected, since galaxies in the SFCG sample are mostly LTGs, and an important part of the light is distributed along the external parts of the galaxy in their star formation processes. This light produced by star formation is bluer, since it is related to the photosphere of young massive stars. If we change to redder filters we will not be receiving that information, since we will be focused on a different wavelength range. Instead, we will be observing the less massive and older stars that are closer to the center, so the radius containing the 50\% of the galaxy light will be smaller. 

When comparing both surveys, on average the {$R_{e}$} is 0.16 {$\pm$} 0.03 arcsec higher in S-PLUS than in DECaLS. Considering the most distant group of the SFCGs sample with a {$z$} = 0.17, the difference would be of 0.48 {$\pm$} 0.09 kpc. In both parameters we find a higher median difference in the \textit{z} filter, which is probably due to the lower S/N in that band in S-PLUS. It is important to remark that this difference is smaller than the usual seeing, so in some cases it even could not be resolved.

\begin{figure}[h!]
    \centering
    \includegraphics[width=\linewidth]{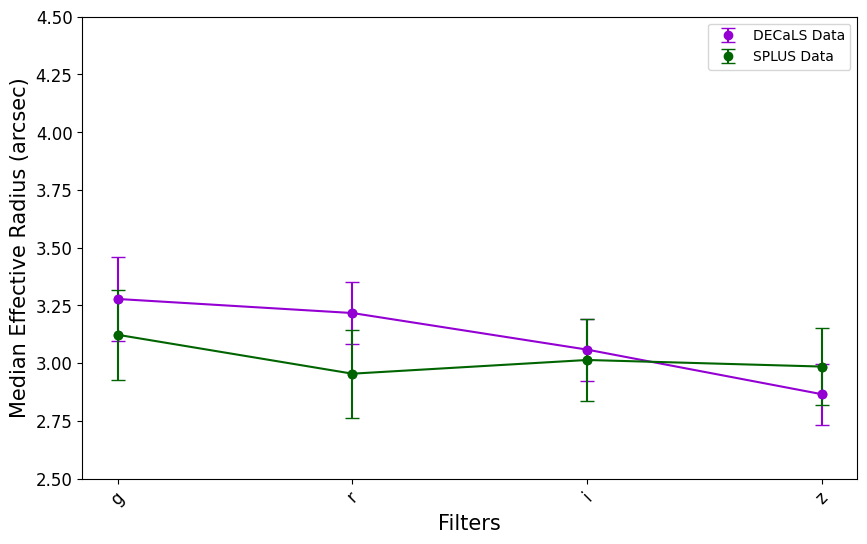}
    \caption{Median effective radius for galaxies in the DECaLS (violet) and S-PLUS (green) surveys, not limiting the sample by magnitude, error bars are obtained using the bootstrapping method, with 1000 re-samples.}
    \label{fig:renolimit}
\end{figure}
    
\begin{figure}[h!]
    \centering
    \includegraphics[width=\linewidth]{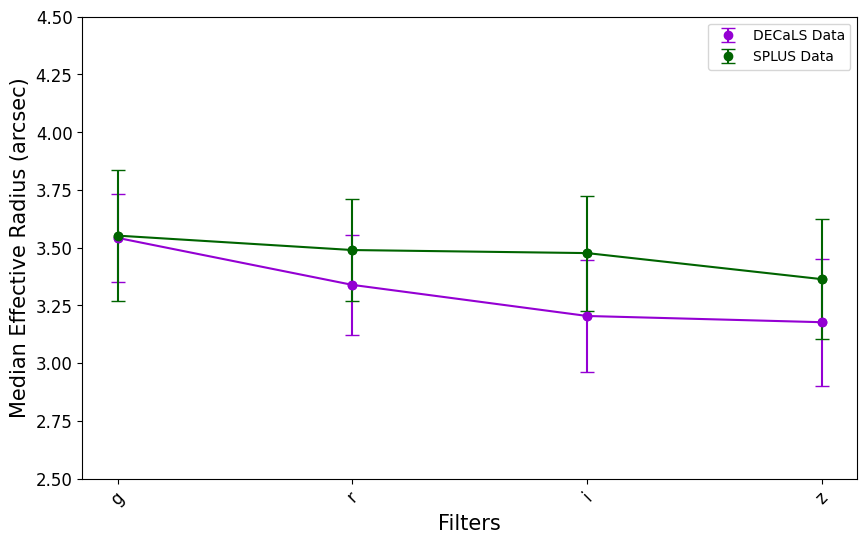}
    \caption{Median effective radius for galaxies in DECaLS (violet) and S-PLUS (green) surveys, limiting magnitudes up to 18.5, error bars are obtained using the bootstrapping method, with 1000 re-samples.}
    \label{fig:relimit}
\end{figure}

\begin{figure}[h!]
    \centering
    \includegraphics[width=\linewidth]{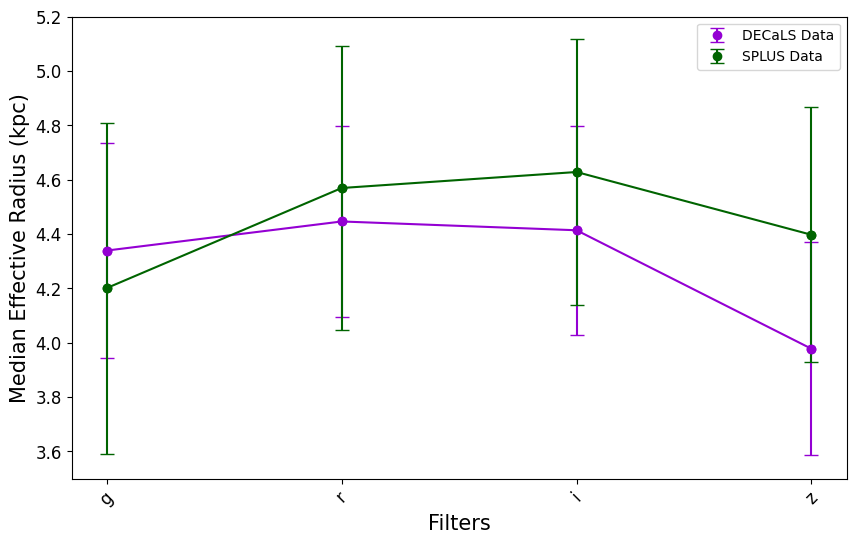}
    \caption{Median effective radius for galaxies in DECaLS (violet) and S-PLUS (green) for galaxies in the SFCGs with at least one member with redshift available (54 galaxies). Error bars are the median uncertainties in each filter computed using bootstrapping with a 68\% confidence interval.}
    \label{fig:median_re_splus_decals_kpc}
\end{figure}

It is important to consider that it is difficult to quantify the real difference in {$R_e$} in arcsec, since depending on the group distance, the distance from arcsec to kpc will be different. Figure \ref{fig:median_re_splus_decals_kpc} shows the median {$R_e$} for galaxies in the DECaLS and S-PLUS survey, considering galaxies in groups with at least one member with redshift available. We observe a mean difference in all filters of $0.22 \pm 0.07$ (kpc), with the highest median difference of $0.46 \pm 0.08$ (kpc) in the z filter.

\begin{figure}[h!]
    \centering
    \includegraphics[width=0.85\linewidth]{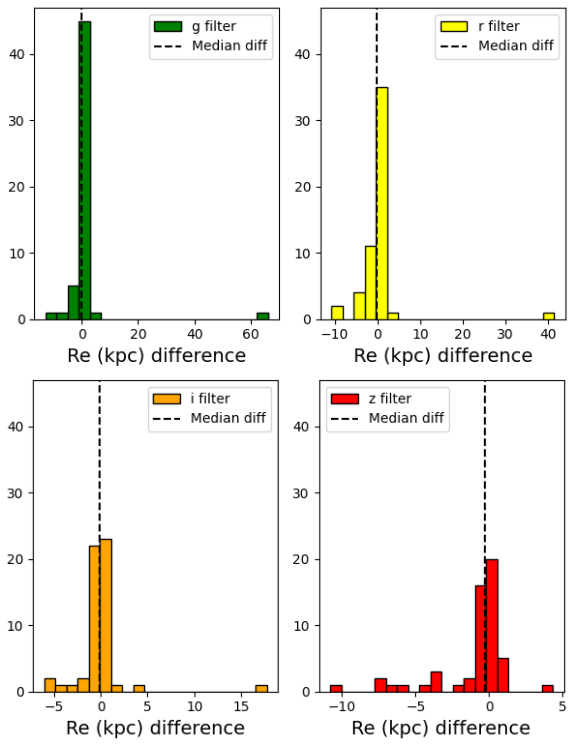}
    \caption{Distribution of the {$R_e$} differences in kpc in each filter, considering the same galaxies as Fig. \ref{fig:median_re_splus_decals_kpc}.}
    \label{fig:diff_re_dist}
\end{figure}

\begin{figure}[h!]
    \centering
    \includegraphics[width=0.85\linewidth]{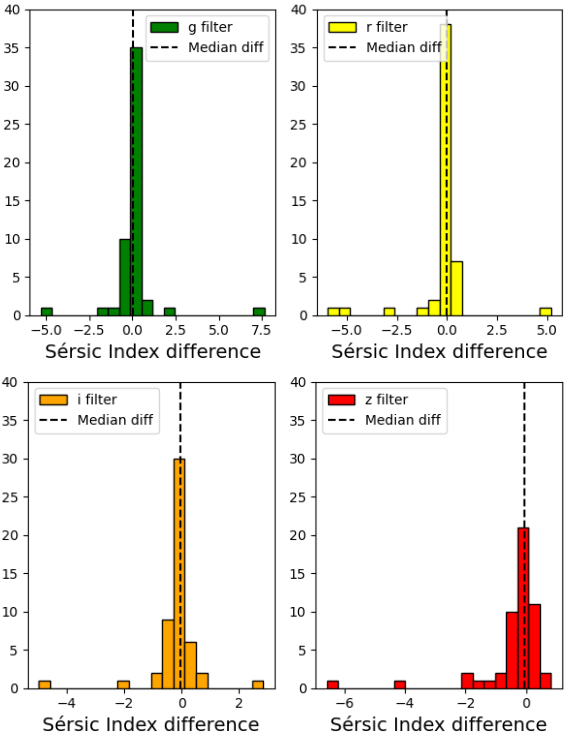}
    \caption{Distribution of the {$n$} differences in each filter, considering the same galaxies as Fig. \ref{fig:median_re_splus_decals_kpc}.}
    \label{fig:diff_n_dist}
\end{figure}

Figures \ref{fig:diff_re_dist} and \ref{fig:diff_n_dist} show the distribution of the differences in {$R_e$} and {$n$} respectively, between each survey for the same galaxies considered in Fig. \ref{fig:median_re_splus_decals_kpc}. We can see that the majority of differences concentrate around 0, median {$R_e$} differences are 0.018, 0.146, 0.101 and 0.260 (kpc) from \textit{g} to \textit{z} filter, with S-PLUS presenting higher parameter values than DECaLS. The same is possible to extract from median {$n$} differences, which are 0.014, 0.017, 0.043 and 0.069 from \textit{g} to \textit{z} filter, being the S-PLUS values higher than DECaLS in all bands but \textit{g} filter.

In both structural parameters we find that the S-PLUS values are slightly higher than in DECaLS, but the differences are not enough to make different interpretation of the results, and since uncertainties overlap, we consider the differences as negligible.

\FloatBarrier
\section{Physical properties and asymmetry distribution for NPC galaxies}\label{sec:appendix_npc}

In this section we provide information about physical properties of galaxies classified through non-parametric methods, as well as the asymmetry distribution of galaxies regarding their type.

\begin{figure}[h!]
    \centering
    \includegraphics[width=0.77\linewidth]{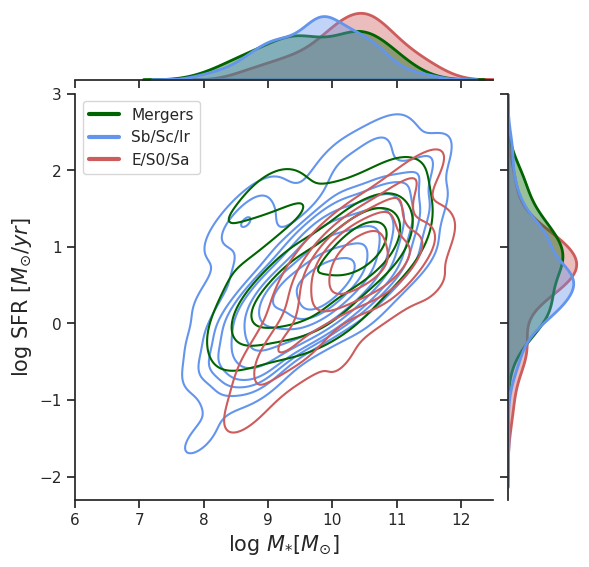}
    \caption{Star formation rate vs. stellar mass for each galaxy type in the NPC, KDE contours represent Sb/Sc/Ir (blue), E/S0/Sa (red), and merger (green) galaxies. The marginal plots show the KDE for each parameter distribution.}
    \label{fig:sfr-mass-no-param}
\end{figure}

\begin{figure}[h!]
    \centering
    \begin{subfigure}
        \centering
        \includegraphics[width=0.9\linewidth]{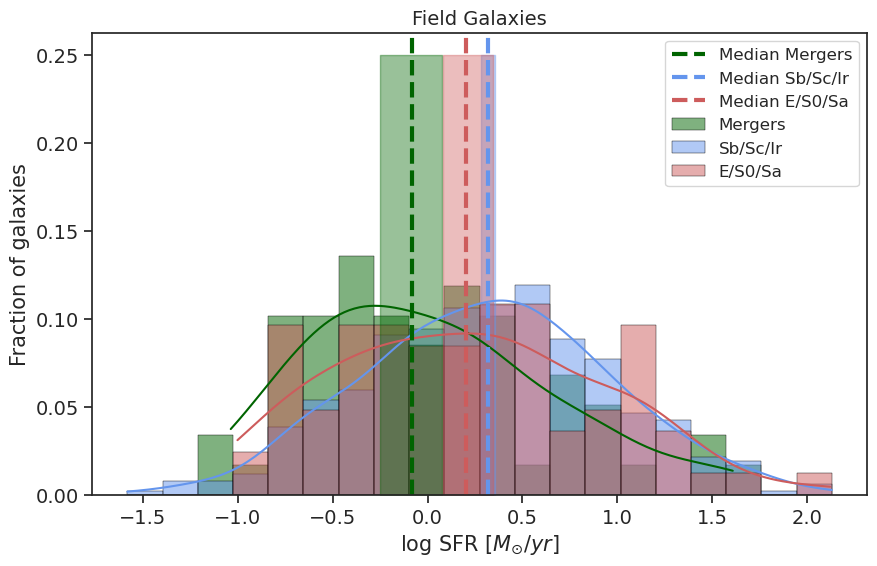}
    \end{subfigure}
    \hfill
    \begin{subfigure}
        \centering
        \includegraphics[width=0.9\linewidth]{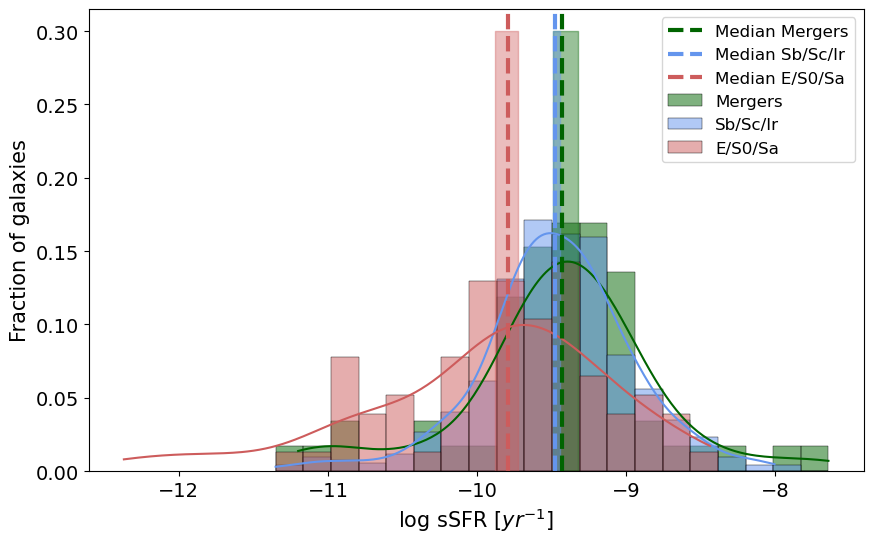}
    \end{subfigure}
    \caption{Star formation rate (top panel) and sSFR (lower panel) for galaxies in each NPC, Sb/Sc/Ir (blue), merger (green), and E/S0/Sa (red) type galaxies in the field. Dashed lines represent the median SFR for each classification, while the shaded regions represent the errors obtained using bootstrapping to the 68\% confidence interval.}
    \label{fig:sfr_ssfr_no_param_cs}
\end{figure}

\begin{figure}[h!]
    \centering
    \includegraphics[width=1\linewidth]{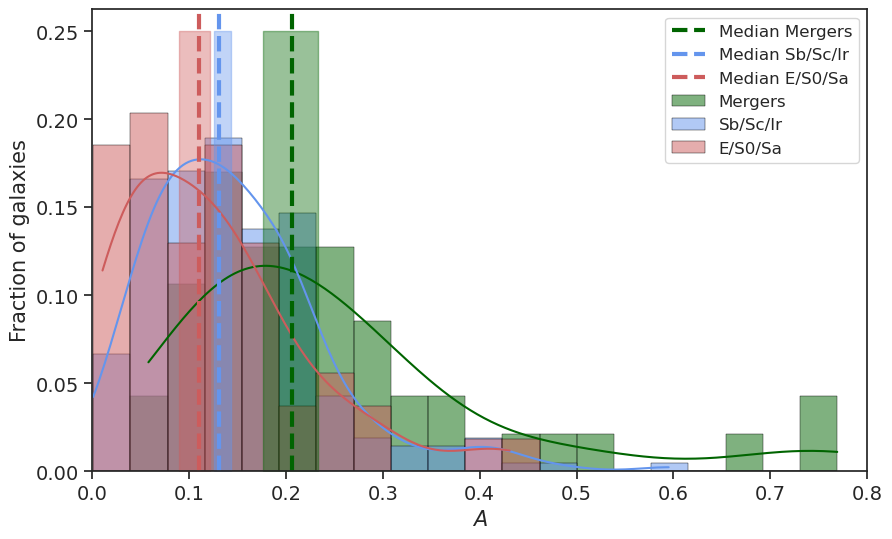}
    \caption{Asymmetry distribution for galaxies in the SFCGs for Sb/Sc/Ir (blue), E/S0/Sa (red), and merger (green) type galaxies. We find that asymmetry is significantly larger for merger galaxies.}
    \label{fig:asym_distribution}
\end{figure}

\end{appendix}

\end{document}